\documentclass[aps,prd,superscriptaddress,floatfix,nofootinbib,notitlepage,12pt]{revtex4-1}

\usepackage{aas_macros}
\usepackage{graphicx}
\usepackage{multirow}
\usepackage{color}
\usepackage{slashed}
\usepackage{comment}
\usepackage{ulem}
\usepackage{soul}
\setstcolor{red}
\usepackage{amsmath,amssymb,amsfonts}
\usepackage{verbatim}
\usepackage{bm}
\usepackage{color}
\usepackage{ulem}
\usepackage{mathtools}
\usepackage{colortbl}
\usepackage[dvipsnames]{xcolor}
\usepackage{slashed}
\usepackage{braket}
\usepackage{adjustbox}
\usepackage{cancel}

\usepackage{ulem}
\usepackage{soul}
\setstcolor{red}

\usepackage[colorlinks]{hyperref}
\hypersetup{
	colorlinks=true,
	linkcolor=blue,
	filecolor=blue,      
	urlcolor=blue,
	citecolor=magenta,
	pdfpagemode=FullScreen
} 

\newcommand{\orcid}[1]{\hspace{1mm}\href{https://orcid.org/#1}{\includegraphics[height=0.3cm,keepaspectratio]{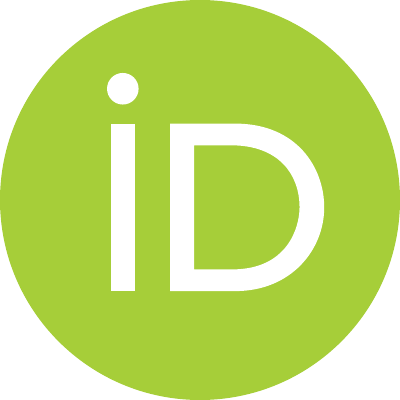}}}

\newcommand{\hc}{\text{h.c}}

\begin{document}
	
\title{Comparison between $\mu^{-}\mu^{+}$ and $e^{-}e^{+}$ colliders for charged Higgs production in 2HDM}	
\author{Brahim Ait Ouazghour\orcid{0009-0006-1419-969X}}
\email{b.ouazghour@gmail.com}
\affiliation{LPHEA, Physics Department, FSSM, Cadi Ayyad University, P.O.B. 2390 Marrakech, Morocco}
\author{Abdesslam Arhrib\orcid{0000-0001-5619-7189}}
\email{aarhrib@gmail.com}
\affiliation{Abdelmalek Essaadi University, FST Tanger B.P. 416, Morocco}
\affiliation{Department of Physics and CTC, National Tsing Hua University, Hsinchu, Taiwan 300}
\author{Kingman Cheung\orcid{0000-0003-2176-4053}}
\email{cheung@phys.nthu.edu.tw}
\affiliation{Department of Physics and CTC, National Tsing Hua University, Hsinchu, Taiwan 300}
\affiliation{Division of Quantum Phases and Devices, School oproduction f Physics,
Konkuk University, Seoul 143-701, Republic of Korea}
\author{Es-said Ghourmin\orcid{0009-0007-1597-8537}}
\email{s.ghourmin123@gmail.com}
\affiliation{Laboratory of Theoretical and High Energy Physics (LPTHE), Faculty of Science, Ibnou Zohr University, B.P 8106, Agadir, Morocco}
\author{Larbi Rahili\orcid{0000-0002-1164-1095}}
\email{rahililarbi@gmail.com}
\affiliation{Laboratory of Theoretical and High Energy Physics (LPTHE), Faculty of Science, Ibnou Zohr University, B.P 8106, Agadir, Morocco}	
\date{\today}
\begin{abstract}
We study the phenomenology of the charged Higgs boson at future muon colliders. We investigate both the pair production $\mu^+ \mu^- \to H^+ H^-$, the single production $\mu^+ \mu^- \to W^\pm H^\mp$, as well as the Vector Boson Fusion (VBF) $\{e^+e^-, \mu^+ \mu^-\} \to \nu \bar{\nu} H^+ H^-$. We show that the neutral Higgs exchange diagrams in the muon collider case can lead to a significant boost in the cross sections  through their Yukawa couplings. Our results for the muon collider are systematically compared to the corresponding ones at $e^+e^-$ machines. It is demonstrated that the vector boson fusion (VBF) $e^+e^- \to \nu \bar{\nu} H^+ H^-$ can compete with the mentioned $2\to 2$ processes. We select benchmark points and perform signal-background analyses, taking into account detector simulations. We demonstrate the discovery region at $5\sigma$ and the excluded region at $2\sigma$ levels at a  3 TeV muon collider.
\end{abstract}

\maketitle

\newpage

\section{Introduction}

The Standard Model (SM) spectrum was completed with the discovery of a
scalar particle that exhibits properties similar to those of the SM Higgs boson by ATLAS \cite{ATLAS:2012yve} and CMS \cite{CMS:2012qbp} in 2012. The majority of measurements regarding the Higgs couplings and cross sections conducted at the LHC are in good agreement with the theoretical framework of the SM \cite{ATLAS:2022vkf, CMS:2022dwd}. So on, with the new LHC run and its projected High Luminosity option \cite{CERN-ACC-2014-0300}, we are now entering an era of precise measurement programs at the LHC to scrutinize the SM.

Despite its success in explaining various phenomena and its predictions aligning with experimental results, however, the SM still exhibits weaknesses, including challenges related to dark matter and dark energy, matter-antimatter asymmetry, the hierarchy problem, and neutrino mass generation. These weaknesses suggest that the SM is merely a low-energy effective theory of a more fundamental one that is yet to be discovered. Several beyond-the-SM (BSM) theories near the TeV scale \cite{Ait-Ouazghour:2020slc,Grzadkowski:2011jks,karahan2014effects, darvishi2018implication,Ouazghour:2018mld} naturally incorporate an extended Higgs sector.

Among the many extensions of BSM theories, the Two-Higgs-Doublet Model (2HDM) is notable for its simplicity, positioning it as a candidate for a BSM model. It can address certain limitations of the Standard Model, such as providing a mechanism for CP violation and explaining the matter-antimatter asymmetry in the universe. The 2HDM is also featured in several high-energy (UV) theories motivated by naturalness and dark matter considerations, like the Minimal Supersymmetric Standard Model (MSSM). In view of this, an additional $SU(2)_{L}$ Higgs doublet is added and the Higgs spectrum is indeed widened compared to the Standard Model (SM). This results in a total of five physical Higgs particles \cite{Branco:2011iw}, including a pair of singly-charged Higgs bosons ($H^\pm$); the latter can be abundantly produced at both hadron and $e^+e^-$ colliders.  

At hadron colliders, the charged Higgs boson can be produced through several channels: from the top decay when the charged-Higgs mass is less than $m_t - m_b$, associated production with a top quark through gluon-gluon fusion or gluon-bottom quark annihilation, and a number of other processes (for a review see Ref.\cite{Akeroyd:2016ymd}). Meanwhile, at $e^+e^-$ colliders, the di-charged Higgs boson is mainly produced through the $s$-channel process $e^+e^-\to \gamma^*, Z^*\to H^+H^-$ \cite{Komamiya:1988rs}. Consequently, its rate depends solely on the charged Higgs boson mass and gauge couplings \cite{Arhrib:1998gr,Guasch:2001hk}. The contribution from $s$-channel diagrams involving neutral Higgs exchange is proportional to the electron mass $(m_e)$ and is thus significantly suppressed. Similarly, the associate production $e^+ e^- \to W^\pm
H^\mp$ in the 2HDM \cite{Arhrib:1999rg,Kanemura:1999tg} and in the
MSSM \cite{Heinemeyer:2016wey,Brein:2006xda,Logan:2002it} is only
mediated at the one-loop level, with the tree-level cross section
being suppressed by $m_e^2$ \cite{Arhrib:1999rg,Kanemura:1999tg,
  Heinemeyer:2016wey,Brein:2006xda, Logan:2002it}. Finally, associated pairwise production of di-charged Higgs boson with $\nu_e \bar{\nu}_e$ is still possible at $e^+e^-$ colliders, and depends on the gauge couplings of the charged Higgs as we will see further on in this study.

Therefore, the discovery of non-standard Higgs bosons at colliders, most notably at the LHC (see Ref.\cite{Kling:2020hmi}), would validate one or more models beyond the SM (BSM). The absence of such a discovery so far has led to current bounds on the masses and couplings of the non-SM Higgs bosons. The High Luminosity LHC (HL-LHC) will improve some of the aforementioned measurements and may find hints of the presence of new physics. However, to pursue such a precise measurement program that was initiated at the LHC, there is a consensus or need
to build a clean environment electron-positron Higgs factory \cite{Abe:2001grn,LinearColliderAmericanWorkingGroup:2001rzk,LinearColliderAmericanWorkingGroup:2001tzv}, which allows detailed studies of the novel SM-like Higgs boson.
%maybe find new particles.
Several projects for $e^+e^-$ machines are
planned, including the Circular Electron Positron Collider (CEPC)
\cite{An:2018dwb}, the Compact Linear Collider (CLIC)
\cite{CLICPhysicsWorkingGroup:2004qvu,Aicheler:2012bya}, the Future
Circular Collider (FCC-ee)
\cite{FCC:2018evy,TLEPDesignStudyWorkingGroup:2013myl}, and the
International Linear Collider (ILC)
\cite{LCCPhysicsWorkingGroup:2019fvj,Moortgat-Pick:2015lbx,Bhattacharya:2023mjr}.

The muon collider (MuC) concept has been around since the 1960s. However,
there have been renewed interests in muon colliders
operating at high energies in the range of multi-TeV
\cite{Delahaye:2019omf,Han:2020uid,Long:2020wfp} in recent years.
There exist a number of studies that
suggested the possibility of using the muon collider to detect
electroweak dark matter \cite{Han:2020uak} and to discover heavy
particles from BSM physics \cite{Costantini:2020stv,Han:2021udl,Vignaroli:2023rxr}.
The primary reasons for building a muon collider are that  it
offers significant physics opportunities to open an
unprecedented new energy frontier for new physics and to provide
a clean leptonic-collision environment for precision studies
~\cite{Capdevilla:2021rwo,Liu:2021jyc,Huang:2021nkl,Yin:2020afe,Buttazzo:2020eyl,Capdevilla:2020qel,Han:2020pif,Han:2020uak,Costantini:2020stv}.

Charged-Higgs pair production and its associated production with a $W$
gauge boson at the future muon collider have been studied in the
context of the Minimal Supersymmetric Standard Model (MSSM)
\cite{Akeroyd:1999xf,Hashemi:2012nz} and of the 2HDM with CP-violation
\cite{Akeroyd:2000zs}. In this work, we investigate charged Higgs
boson production at the future muon collider in the framework of the
2HDM with different Yukawa textures, by investigating the following $2\to2$
processes: $\mu^+ \mu^- \to H^\pm H^\mp$ and $\mu^+ \mu^- \to W^\pm
H^\mp$.Both processes have additional contributions from the $s$-channel neutral-Higgs exchange and the $t$-channel neutrino exchange due to the large Yukawa of the muon. These new contributions may enhance/suppress the cross section at the muon collider with respect to the $e^+e^-$ case, given the fact that the muon mass is about 207 times larger than the electron mass.
Furthermore, the presence of neutral-Higgs exchange in the
$s$-channel for both processes may offer the possibility of resonance
enhancement. Additionally, the process $\mu^+ \mu^- \to W^\pm H^\mp$ offers the
possibility of searching for the charged Higgs with mass up
to $\sqrt{s}-m_W$, in contrast to
$\mu^+ \mu^- \to H^\pm H^\mp$, which only probes up to
$m_{H^\pm}<\sqrt{s}/2$.  We will also show that  the vector boson fusion (VBF) $e^+e^- \to \nu \bar{\nu} H^+ H^-$ 
can compete with the aforementioned $2\to 2$ processes.

We present results for different Yukawa textures of the 2HDM. The 2HDM Types II and X are of particular interest, as the neutral-Higgs coupling to a pair of muons experiences enhancement in the large $\tan\beta$ limit and may increase the cross section accordingly. Our numerical results are presented after scrutinizing the
2HDM parameter space by imposing various theoretical constraints (unitarity, perturbativity, and vacuum stability) and experimental ones (from SM-like Higgs boson discovery data, BSM Higgs boson exclusions data, Electroweak precision tests (EWPT) and flavor physics). In the allowed parameter space, we then calculate the signal and various SM backgrounds, followed by a full Monte Carlo (MC) analysis and estimations of the sensitivity at the center-of-mass energy $\sqrt{s}= 3$ TeV.

The paper is structured as follows: we begin with a brief review of
the 2HDM in the following section, discussing the scalar sector, the various couplings required, and the relevant theoretical and experimental constraints.
Then, we present the details of the calculations for $\mu^+ \mu^- \to H^\pm H^\mp$ and $\mu^+ \mu^- \to W^\pm H^\mp$ in Sec.\ref{section3}. Section~\ref{Section4} presents the numerical results of our study taking into account theoretical constraints, flavor-physics
constraints as well as experimental ones from LEP-II, Tevatron and LHC. In Sec.\ref{section5}, we detail the Monte Carlo analysis and calculate the significance for the charged-Higgs discovery at the  3 TeV muon collider. Lastly,  we provide our concluding remarks in Sec.\ref{sec:conclusion}.

\section{2HDM review, theoretical and experimental constraints}
\label{sec:model}
\subsection{2HDM review}
\label{subsec:review}
We briefly discuss the basic features of the  two-Higgs-doublet model
(2HDM) \cite{Lee:1973iz,Branco:2011iw} 
and the various Yukawa textures \cite{Paschos:1976ay,Glashow:1976nt}.
In the 2HDM, in addition to SM doublet $\Phi_1$, a new doublet
$\Phi_2$ with a hypercharge $+1$ is added to the Higgs sector,
where we assume that CP is not spontaneously broken.
The two Higgs-scalar doublets can be parametrized by :
\begin{equation}
\Phi_1 = \left(
\begin{array}{c}
\phi_1^+ \\
\phi_1^0 \\
\end{array}
\right)
\quad {\rm and}\quad 
\Phi_2 = \left(
\begin{array}{c}
\phi_2^+ \\
\phi_2^0 \\
\end{array}
\right)
\end{equation}
with $\phi_i^0 = (v_i+\psi_i+ i \eta_i)/\sqrt{2}$, $i=1,2$. 
The general scalar potential $SU(2)_L\times U(1)_Y$
invariant under the SM transformations
can be written as \cite{Branco:2011iw}:
\begin{eqnarray}
	V(\Phi_1,\Phi_2) &=& m_{11}^2 \Phi_1^\dagger\Phi_1+m_{22}^2\Phi_2^\dagger\Phi_2-[m_{12}^2\Phi_1^\dagger\Phi_2+{\rm h.c.}] + \frac{\lambda_1}{2}(\Phi_1^\dagger\Phi_1)^2 + \frac{\lambda_2}{2}(\Phi_2^\dagger\Phi_2)^2\nonumber\\
	&+&\lambda_3(\Phi_1^\dagger\Phi_1)(\Phi_2^\dagger\Phi_2)
	+\lambda_4(\Phi_1^\dagger\Phi_2)(\Phi_2^\dagger\Phi_1) 
	+\left\{\frac{\lambda_5}{2}(\Phi_1^\dagger\Phi_2)^2+h.c\right\}  \label{pot1} \;.
	\label{scalar_pot}
\end{eqnarray}
In the above potential, all $m_{11}^2$, $m_{22}^2$, $m_{12}^2$
parameters as well as the $\lambda_{i}\,(i=1,2,3,4,5)$ couplings are
assumed to be real to insure that our potential is CP conserving.  We
also advocate a discrete $Z_2$ symmetry in order to avoid the Flavor
Changing Neutral Currents (FCNC) at tree level. Such a $Z_2$ symmetry
is only softly broken by the bilinear term proportional to $m_{12}^2$
parameter.

After EWSB takes place, the outcome of the 8 degrees of freedom initially
present in the 2 Higgs doublet fields:
3 are taken by the Goldstone bosons to give masses to the gauge bosons
$W^\pm$ and $Z$, and the rest 5 become the five physical Higgs states,
including a pair of charged Higgs bosons, a CP-odd boson $A$, and 2
CP-even bosons: $H$ and $h$ with $m_h < m_H$.
One of the neutral CP-even Higgs bosons would
be identified as the 125 GeV Higgs-like particle observed at the LHC.
The combination $v^2=v_1^2+v_2^2=(2\sqrt{2} G_F)^{-1}=4 m_W^2/g^2$ can
be used to fix one of the vevs as a function of $G_F$ 
and $\tan\beta$, together with the two minimization conditions,
the scalar potential in Eq.(\ref{scalar_pot}) has seven independent
parameters:
\begin{equation}
\label{eq:modelpara}
\alpha,\quad \tan\beta=\frac{v_2}{v_1},\quad  m_{h}\,=125\ {\rm GeV},\quad m_{H},\quad m_A,\quad 
m_{H^\pm}\quad \ \rm{and}\ \  m_{12}^2,
\end{equation}
where $\alpha$ and $\beta$ are, respectively, the CP-even mixing angle and CP-odd mixing angle. In this work, we consider $h$ to be the SM-like boson observed at the LHC with $m_h=125$ GeV, so the scalar potential is completely described by six independent parameters. 

On the other hand, it is well known in the Yukawa sector
that if we assume that both Higgs doublets couple to all fermions,
we would end up with large tree-level FCNC's mediated by the
neutral Higgs bosons. 
In order to avoid such large FCNCs, the 2HDM needs to satisfy the
Paschos-Glashow-Weinberg theorem \cite{Paschos:1976ay,Glashow:1976nt},
which states that all fermions with the same quantum 
numbers can couple to the same Higgs doublet to avoid
tree-level FCNC's.
One can then have 4 different types (I, II, X, Y) of Yukawa textures. 
In the Type-I model, only the second doublet $\Phi_2$ interacts
with all the fermions,
while in the Type-II model $\Phi_2$ interacts with up-type quarks
and $\Phi_1$ interacts with the charged leptons and down-type quarks.
The Type-X model is where $\Phi_2$ couples to all quarks and $\Phi_1$ couples 
to all leptons,
while in the Type-Y (flipped) model the down-type quarks acquire masses
from their couplings to $\Phi_1$, and the charged leptons and up-type quarks
couple to $\Phi_2$.

In terms of the mass eigenstates of the neutral- and charged-Higgs boson
fields, the Yukawa interactions can be written as:
 \begin{eqnarray}
-{\cal L}_Y &=& \sum_{f=u,d,\ell} \frac{m_f }{v} \left[ \kappa^f_h  \bar f f  h + \kappa^f_H \bar f f H - i \kappa^f_A \bar f \gamma_5 f A \right]  +  \label{eq:Yukawa_CH} \\
&&+ \frac{\sqrt{2}}{v} \left[ \bar u_{i} V_{ij}\left( m_{u_i}  \kappa^{u}_A P_L + \kappa^{d}_A  m_{d_j} P_R \right)d_{j}  H^+ \right]   
+ \frac{\sqrt{2}}{v}  \bar \nu_L  \kappa^\ell_A m_\ell \ell_R H^+ +\hc \;.
\nonumber 
\end{eqnarray} 

\begin{table}
\begin{center}
\begin{tabular}{|c|c|c|c|c|c|c|c|c|c|}
 \hline 
 & $\kappa_h^u$ & $\kappa_h^d$ & $\kappa_h^l$ & $\kappa_H^u$ & $\kappa_H^d$ & $\kappa_H^l$ & $\kappa_A^u$ & $\kappa_A^d$ & $\kappa_A^l$ \\
  \hline 
    Type-I & $c_\alpha/s_\beta$ & $c_\alpha/s_\beta$& $c_\alpha/s_\beta$ & $s_\alpha/s_\beta$ & $s_\alpha/s_\beta$ & $s_\alpha/s_\beta$ & $c_\beta/s_\beta$ & 
    $-c_\beta/s_\beta$ & $-1/\tan\beta$ \\ \hline
    Type-II & $c_\alpha/s_\beta$ & $-s_\alpha/c_\beta$& $-s_\alpha/c_\beta$ & $s_\alpha/s_\beta$ & $c_\alpha/c_\beta$ & $c_\alpha/c_\beta$ & $c_\beta/s_\beta$ & 
    $s_\beta/c_\beta$ & $\tan\beta$ \\ \hline 
    Type-X & $c_\alpha/s_\beta$ & $c_\alpha/s_\beta$& $-s_\alpha/c_\beta$ & $s_\alpha/s_\beta$ & $s_\alpha/s_\beta$ & $c_\alpha/c_\beta$ & $c_\beta/s_\beta$ & 
    $-c_\beta/s_\beta$ & $\tan\beta$ \\ \hline
    Type-Y & $c_\alpha/s_\beta$ & $-s_\alpha/c_\beta$& $c_\alpha/s_\beta$ & $s_\alpha/s_\beta$ & $c_\alpha/c_\beta$ & $s_\alpha/s_\beta$ & $c_\beta/s_\beta$ & 
    $s_\beta/c_\beta$ & $-1/\tan\beta$ \\ \hline 
    \end{tabular}
 \end{center}
 \caption{Yukawa couplings of the $h$, $H$, and $A$ Higgs bosons to the quarks and leptons in 2HDM.} 
\label{coupIII}
\end{table} 

Whilst, the reduced couplings of the lighter Higgs boson, $h$, to
either $WW$ or $ZZ$ are given by $\sin(\beta-\alpha)$, on the other
hand, the couplings of the heavier Higgs boson, $H$, are equivalent to
the SM couplings multiplied by $\cos(\beta-\alpha)$. Notably, the
coupling between the pseudoscalar $A$ and vector bosons is absent
due to CP invariance.

Since throughout this study a few couplings are of crucial importance,
we explicitly list them here:
the following identities,
\begin{eqnarray}
&& \kappa_{h}^{\ell} (\text{II,X}))=-\frac{\sin\alpha}{\cos\beta}=s_{\beta-\alpha}-\tan\beta\,c_{\beta-\alpha} \nonumber \\
&& \kappa_{H}^{\ell} (\text{II,X})=\frac{\cos\alpha}{\cos\beta}=c_{\beta-\alpha}+\tan\beta\,s_{\beta-\alpha} \;. \nonumber
\end{eqnarray}
It is clear that $-\frac{\sin\alpha}{\cos\beta}$ and
$\frac{\cos\alpha}{\cos\beta}  $ exhibit some enhancement for
large $\tan\beta$. Note that close to the decoupling limit
$\sin(\beta-\alpha)\approx 1$, which is also favored by LHC data, 
the $h$ couplings to fermions reduce to unity.

For completeness, we also list the Feynman rules for pure scalar interactions :
\begin{eqnarray}
&&g_{h H^+H^-} =  -\frac{1}{v}\bigg[ \Big( 2m_{H^\pm}^2-m_h^2 \Big) s_{\beta-\alpha} + \Big(m_h^2 - 2\frac{m_{12}^2}{s_{2\beta}} \Big)\frac{c_{\beta+\alpha}}{s_\beta c_\beta} \bigg] \nonumber\\
&& g_{H H^+H^-}  =  -\frac{1}{v}\bigg[ \Big( 2m_{H^\pm}^2-m_H^2 \Big) c_{\beta-\alpha} +\Big(m_H^2 - 2\frac{m_{12}^2}{s_{2\beta}} \Big)\frac{s_{\beta+\alpha}}{s_\beta c_\beta} \bigg] \nonumber\\
&&g_{Hhh} = -\frac{c_{\beta-\alpha}}{v s_{2\beta}^2}\bigg[  (2 m_h^2 + m_H^2) s_{2\alpha} s_{2\beta}   - 2 m_{12}^2 
(3  s_{2\alpha}  - s_{2\beta} )\bigg] \nonumber\\
&&g_{h H^+G^-} =  -\frac{c_{\beta-\alpha}}{v}(m_h^2 - m_{H^\pm}^2) \nonumber\\
&&g_{H H^+G^-}=   -\frac{s_{\beta-\alpha}}{v}(m_H^2 - m_{H^\pm}^2)\label{thdm:h1hphm}
\end{eqnarray}

The relevant part of the Lagrangian describing the interactions 
of the gauge bosons with scalars is:
\begin{eqnarray}
{\cal L} &=& \frac{g}{2}W_{\mu}^+ ( (H^- \stackrel{\leftrightarrow}{\partial}^{\mu} A)-
 ic_{\beta-\alpha} (H^- \stackrel{\leftrightarrow}{\partial}^{\mu} h)+
  i s_{\beta-\alpha} (H^- \stackrel{\leftrightarrow}{\partial}^{\mu} H))+ h.c\nonumber\\
  &&  + \frac{g}{2 c_W }Z_{\mu} ( c_{\beta-\alpha}  (A \stackrel{\leftrightarrow}{\partial}^{\mu} h)-
 s_{\beta-\alpha}  (A \stackrel{\leftrightarrow}{\partial}^{\mu} H)) \nonumber\\
&&+ ( ie\gamma_\mu+ i\frac{g (c_W^2-s_W^2)}{2 c_W} Z_{\mu}^+)  (H^\mp\stackrel{\leftrightarrow}{\partial}^{\mu} H^\pm ) 
\label{lag}
\end{eqnarray}
where $\{s_W,c_W\}=\{\sin\theta_W, \cos\theta_W\}$ and $\theta_W$ stands for the Weinberg angle.

\subsection{Theoretical and Experimental Constraints}
\label{constraint}
There are both theoretical and experimental constraints that must be
satisfied by the parameter space of the 2HDM. Theoretical constraints
due to theoretical consistency conditions include vacuum stability,
perturbative unitarity, and perturbativity.
The experimental constraints include  the measurements of Higgs boson
properties, flavor-changing neutral current observables, and electroweak
precision observables.  These constraints play an important role in
determining the allowed regions of the 2HDM parameter space and can
guide the searches for new physics beyond the SM. So, we will
provide a brief description of these constraints in this subsection.
\begin{itemize}
\item \textbf{Perturbative unitarity}: 
New physics beyond the SM must still obey the fundamental 
principles such as perturbative unitarity. 
Hence, in order to be consistent with such requirements
within the 2HDM, one can impose perturbative unitarity in a
variety of scattering processes among 
the various scalars and gauge bosons. These constraints have been
taken from Refs~\cite{Kanemura:1993hm,Akeroyd:2000wc,Arhrib:2000is}.
\item \textbf{Perturbativity}: To avoid a non-perturbative theory,
  the quartic couplings of the scalar potential must obey the following
  conditions: $|\lambda_i|<8 \pi$ for each $i=1,..,5$ ~\cite{Branco:2011iw}.
\item \textbf{Vacuum stability}: It is an important constraint that ensures
  the scalar potential to be bounded from below when the fields are allowed
  to take on larger and larger values. To fulfill this need,
  the Higgs potential must be positive in any direction of the fields
  $\Phi_i$, and as a consequence the
  conditions \cite{Barroso:2013awa,deshpande1978pattern}, 
\begin{eqnarray}
\lambda_{1,2}>0,  \quad
\lambda_3>- \sqrt{\lambda_1\lambda_2}, \quad
\lambda_3+\lambda_4-|\lambda_5|> - \sqrt{\lambda_1\lambda_2},
\label{eq:VB}
\end{eqnarray}	
must be satisfied in the whole parameter space.
\item[\textbullet]{\bf The EW precision observables}: are being
  utilized to quantify the deviations from the predictions of the
  SM. Those observables, namely $S, T$ and $U$~\cite{Peskin:1991sw},
  should be within 95\% C.L. of their experimental measurements, and
  the current fit values are given by~\cite{Lu:2022bgw}:
\begin{eqnarray}
S= 0.06\pm 0.10,\  \ T = 0.11\pm 0.12,\ \ U = -0.02\pm 0.09,  \nonumber \\
\rho_{ST} = 0.90,\ \ \rho_{SU} = -0.57,\ \ \rho_{TU} = -0.82\;\;\;\;\;\;\;\;\; \nonumber 
\end{eqnarray}
where $\rho_{ST}$ is the correlation parameter.
\paragraph*{}
The constraints mentioned above have been incorporated into
\texttt{2HDMC-1.8.0} \cite{Eriksson:2009ws}, which is publicly
available. This code is utilized to explore the parameter space of the
2HDM and assess its compatibility with the aforementioned constraints,
as well as to calculate the Higgs branching ratios at each
point. Additionally, 2HDMC includes an interface to
\texttt{HiggsBounds-5.10.1} \cite{Bechtle:2020pkv,Bechtle:2015pma} and
\texttt{HiggsSignals-2.6.1} \cite{Bechtle:2020uwn}, as described
below.
\item {\bf BSM Higgs boson exclusions}: To make our study be compatible
  with the existing exclusion limits at the 95\% confidence level from
  Higgs searches at LEP, LHC, and Tevatron, we use
  the \texttt{HiggsBounds-5.10.1}  \cite{Bechtle:2020pkv,Bechtle:2015pma}.
  The primary search channels that have implications for the Type-II and
  Type-X 2HDM include:
\begin{itemize}
\item Type-II
\begin{itemize}
\item $p p \rightarrow A \rightarrow Z Z \rightarrow l^+ l^+ l^- l^-,\;l^+l^-qq,\;l^+l^- \nu \bar{\nu} $ \cite{CMS:2018amk}
\item $b b \rightarrow A \rightarrow Z h \rightarrow l^+ l^- b b $ \cite{CMS:2019qcx}
\item $p p \rightarrow A \rightarrow \tau^+ \tau^- $ \cite{ATLAS:2020zms}
\item $p p \rightarrow H \rightarrow \tau^+ \tau^- $ \cite{ATLAS:2020zms}
\item $p p \rightarrow H \rightarrow Z Z \rightarrow l^+l^-l^+l^-, l^+l^-qq,l^+l^-\nu\bar{\nu} $\cite{CMS:2018amk}
\item $p p \rightarrow H \rightarrow V V $\cite{ATLAS:2018sbw}
\item $p p \rightarrow A \rightarrow H Z \rightarrow b\bar{b}l^+l^- $\cite{ATLAS:2018oht}
\item $p p \rightarrow H \rightarrow A Z \rightarrow b\bar{b}l^+l^- $\cite{ATLAS:2018oht}
\item $ g g \rightarrow A \rightarrow h Z \rightarrow b\bar{b}l^+l^- $\cite{ATLAS:2020pgp}
\end{itemize}
\item Type-X \\
\begin{itemize}
	\item $ g g \rightarrow A \rightarrow h Z \rightarrow b\bar{b}l^+l^- $\cite{ATLAS:2020pgp}
	\item $p p \rightarrow h \rightarrow \tau^+ \tau^- $ \cite{CMS:2012jmc}
	\item $p p \rightarrow H \rightarrow Z Z \rightarrow l^+l^-l^+l^-, l^+l^-qq,l^+l^-\nu\bar{\nu} $\cite{CMS:2018amk}
	\item $ g g \rightarrow A \rightarrow h Z \rightarrow b\bar{b}\tau^+ \tau^- $\cite{ATLAS:2015kpj}
	\item $p p \rightarrow A \rightarrow \tau^+ \tau^- $ \cite{ATLAS:2020zms}
	\item $t \rightarrow H^{+} b \rightarrow \tau^+ \nu b$\cite{ATLAS:2018gfm,CMS:2015lsf}
	\item $p p \rightarrow A \rightarrow \tau^+ \tau^- $ \cite{CMS:2015mca,CMS:2018rmh}
	\item $ g g \rightarrow A \rightarrow H Z \rightarrow \tau^+ \tau^- l^+l^- $\cite{CMS:2016xnc}
	\item $p p \rightarrow H \rightarrow \tau^+ \tau^- $ \cite{CMS:2015mca,ATLAS:2020zms}
	\item $e^+ e^- \rightarrow H^+ H^- \rightarrow 4q,\tau^+ \nu \tau^- \bar{\nu} $ \cite{ALEPH:2013htx}
	\item $ g g \rightarrow A \rightarrow h Z \rightarrow b\bar{b}l^+l^- $\cite{CMS:2019qcx}
	\item $ p p \rightarrow A \rightarrow H Z \rightarrow b\bar{b}l^+l^- $\cite{ATLAS:2018oht}
	\item $p p \rightarrow H^{+} t \bar{b} \rightarrow \tau^+ \nu t \bar{b}$\cite{ATLAS:2018gfm}
	\item $ p p \rightarrow H \rightarrow hh \rightarrow \gamma \gamma b\bar{b} $\cite{ATLAS:2018dpp}
	\item $p p \rightarrow H^{+} \bar{t} b \rightarrow t \bar{b} \bar{t} b$\cite{ATLAS:2020jqj}
	\item $p p \rightarrow h \rightarrow \tau^+ \tau^- $ \cite{CMS:2015mca,CMS:2018rmh}
	\item $p p \rightarrow g g \rightarrow A  \rightarrow \tau^+ \tau^- $ \cite{ATLAS:2014vhc}
	\item $ p p \rightarrow H \rightarrow h Z \rightarrow \tau^+ \tau^- l^+l^- $\cite{ATLAS:2015kpj}
	\item $ p p\rightarrow A /VBF/WA /ZA /ttA \rightarrow \gamma \gamma$ \cite{ATLAS:2017ayi}
	\item $ p p \rightarrow H \rightarrow h h \rightarrow b\bar{b}b\bar{b} $\cite{ATLAS:2018rnh}
	\item $ p p \rightarrow H \rightarrow h h \rightarrow b\bar{b}/ \tau^+ \tau^- /W^+ W^-/\gamma \gamma $\cite{ATLAS:2019qdc}
	\item $t \rightarrow H^{+} b \rightarrow c \bar{b} b$\cite{CMS:2018dzl}
	\item $ g g \rightarrow H \rightarrow t\bar{t} $\cite{CMS:2019pzc}
\end{itemize}
\end{itemize}

\item {\bf SM-like Higgs boson properties}:
  In the same vein, \texttt{HiggsSignals-2.6.1} \cite{Bechtle:2020uwn}
  is employed to check the compatibility of the SM-like scalar boson
  with the Higgs signal rate constraints from various searches and
  take into account the recent LHC 13 TeV results.
\begin{table}[!hb]
\centering
\setlength{\tabcolsep}{7pt}
\renewcommand{\arraystretch}{1.2} %
\begin{tabular}{|l||c|c|}
\hline
Observable&Experimental result&95\% C.L. Bounds\\\hline
BR($B_{\mu}\to \tau\nu$)\cite{Haller:2018nnx}&$(1.06 \pm 0.19) \times 10^{-4}$&$ [0.68\times 10^{-4} , 1.44\times 10^{-4} ]$\\\hline
BR($B_{s}^{0}\to \mu^{+}\mu^{-}$)\cite{Haller:2018nnx}&$(2.8 \pm 0.7) \times 10^{-9}$&$[1.4 \times 10^{-9}, 4.2\times 10^{-9}]$\\\hline
BR($B_{d}^{0}\to \mu^{+}\mu^{-}$)\cite{Mahmoudi:2008tp}&$(3.9\pm 1.5)\times10^{-10}$&$[0.9\times 10^{-10}, 6.9\times10^{-9}$\\\hline
BR($\bar{B}\to X_{s}\gamma$)\cite{Haller:2018nnx,HFLAV:2016hnz}&$(3.32\pm 0.15)\times10^{-4}$&$[3.02\times 10^{-4} , 3.61\times 10^{-4}]$\\\hline
\end{tabular}
\caption{
Experimental results of 
  $B_{\mu}\to \tau\nu$, $B_{s,d}^{0}\to \mu^{+}\mu^{-}$ and $\bar{B}\to X_{s}\gamma$ at 95$\%$ C.L.}
\label{Tab2}
\end{table}
\item[\textbullet]{\bf Flavor constraints}: Related to the the B-physics
  observables, and for this purpose we used the \textbf{Superiso v4.1}
  tool \cite{Mahmoudi:2008tp} to calculate the relevant results.
  We then perform consistency checks at a 2$\sigma$ C.L., taking into
  account the available experimental measurements as reported in
  Table.\ref{Tab2}. 
\end{itemize} 

\section{Computational Procedure steps}
\label{section3}
In this section, we list all processes under investigation in this study. 
We give the contributing tree-level Feynman diagrams, their corresponding amplitudes and the corresponding squares of amplitudes. 
We use the Mathematica packages: \texttt{FeynArts} \cite{Hahn:2000kx} and
\texttt{FormCalc}\cite{Hahn:1998yk}
to generate the amplitudes and  to compute the corresponding cross sections.
We also did a check of our calculation with FormCalc outputs and
find perfect agreement.

\subsection{Charged Higgs pair production}
\label{section3subsec1}
\begin{figure}[!ht]
\centering
\includegraphics[width=0.75\textwidth]{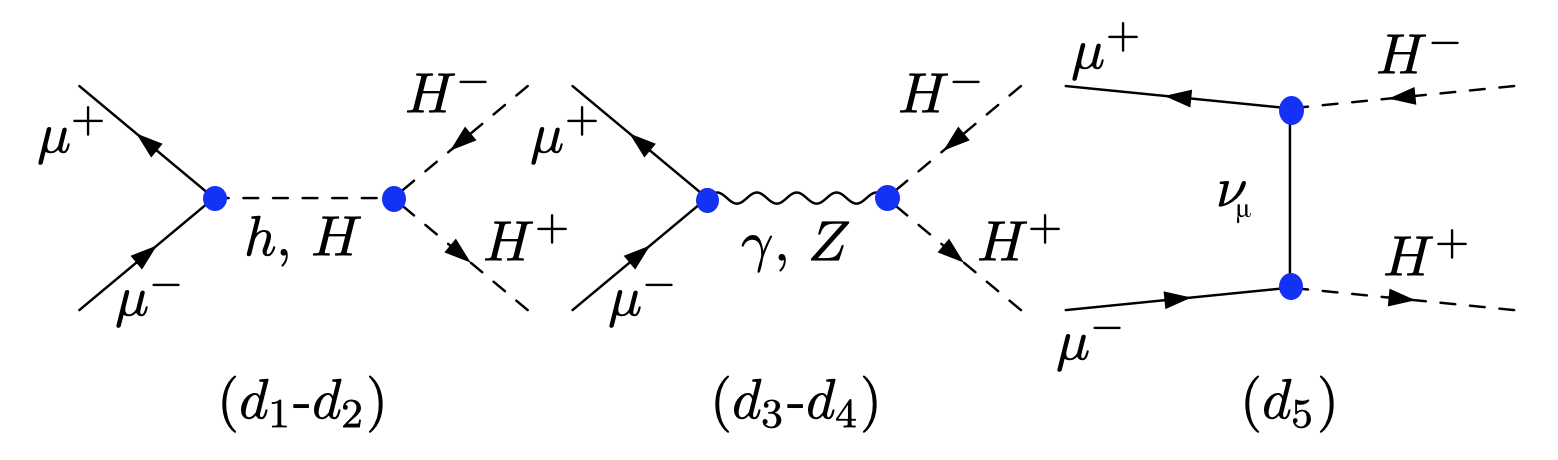}%
\caption{Tree-level Feynman diagrams for $H^{+}H^{-}$ at muon collider in the 2HDM}
\label{diag:2mu2HpHm}
\end{figure}

The tree-level Feynman diagrams contributing to $\mu^{+} \mu^{-} \rightarrow H^{+}H^{-} $ in the 2HDM are given in the Fig.\ref{diag:2mu2HpHm}. 

\noindent
At tree level, the 
conventional Drell--Yan mechanism $\mu^+\mu^-\to \gamma^*,Z^* \to H^+H^-$
in Fig. \ref{diag:2mu2HpHm}$(d_{3,4})$, 
the $s$-channel neutral-Higgs exchange
$\mu^+\mu^-\to h^*, H^* \to H^+H^-$ in Fig. \ref{diag:2mu2HpHm}$(d_{1,2})$,
and the neutrino-exchange diagram in the $t$  channel in
Fig.\ref{diag:2mu2HpHm}$(d_{5})$  all contribute to the process. 
The $s$-channel Higgs diagrams would be largely enhanced
near the resonance region $\sqrt s \approx m_H$,
while the $t$-channel diagram in Fig.\ref{diag:2mu2HpHm} suffer
from the Yukawa-coupling suppression at two vertices, 
but the large $\tan\beta$ value can give some amplification.
If the Yukawa coupling of the charged Higgs to the muon becomes large,
then it may give some enhancement for large center-of-mass energies. 

We use $p_{1,2}$ for the momenta  of the incoming $\mu^-$, $\mu^+$,
and $k_{1,2}$ for the momenta of the outgoing charged Higgs bosons
$H^\mp$. The  four-momenta are defined in the center-of-mass system as:
\begin{eqnarray}
p_{1,2} & = & \big(\frac{\sqrt{s}}{2},\,0\,,0\,, \pm \frac{\sqrt{s}}{2} \big) \label{2mu2HpHm:p1}\nonumber\\
k_{1,2} & = & \big(\frac{\sqrt{s}}{2},\,\pm \frac{\sqrt{s}}{2}\,\beta_{H}\sin\theta,0,\,\pm \frac{\sqrt{s}}{2}\,\beta_{H}\cos\theta\big)\;, \label{2mu2HpHm:k1}
\end{eqnarray}
where $\beta_{H} = \sqrt{1-4\,m_{H^\pm}^2/s}$, $\sqrt{s}$ is the center-of-mass
energy, and $\theta$ is the scattering angle between $\mu^+$ and $H^+$
in the center-of-mass frame.
The Mandelstam variables $s, t $ and $u$ are defined by:
\begin{eqnarray}
s & = & \big(p_{1}+p_{2}\big)^2 = \big(k_{1}+k_{2}\big)^2 \label{2mu2HpHm:s}\nonumber\\
t & = & \big(p_{1}-k_{1}\big)^2 = \big(p_{2}-k_{2}\big)^2 = m_{H^\pm}^2 -\frac{s}{2} +\frac{s}{2} \beta_H \cos\theta \label{2mu2HpHm:t}\nonumber\\
u & = & \big(p_{1}-k_{2}\big)^2 = \big(p_{2}-k_{1}\big)^2 = m_{H^\pm}^2 -\frac{s}{2} -\frac{s}{2} \beta_H \cos\theta  \label{2mu2HpHm:u} \;.
\end{eqnarray}

Following the Feynman rules, the matrix elements for this process are
given by:
\begin{eqnarray}
M_{0}^{\gamma} &=&    -2\frac{ e^2}{s}   \bar{v}(p_2) \slashed{k}_2 u(p_1) \nonumber \\
% \bar{v}(p_2)(+ie)\gamma ^\mu u(p_1) \frac{(-ie)g_{\mu \nu}}{(k_1+k_1)^2} (k_2 -k_1)_{\nu}\\
M_{0}^{z} &=&  \frac{2g_H e^2}{s-m_Z^2+i m_Z \Gamma_Z} \left[ g_V \bar{v}(p_2)  \slashed{k}_2 u(p_1)  - g_A  \bar{v}(p_2)  \slashed{k}_2 \gamma ^5 u(p_1)  \right]  \nonumber\\
% \bar{v}(p_2)(+ie)\gamma ^\mu g_H(g_V-g_A \gamma ^5) u(p_1) \frac{ i(-e)g_{\mu \nu}}{(k_1+k_1)^2-M_Z^2} (k_2 -k_1)_{\nu} \\
M_{0}^{\nu} &=& - \frac{g^2 m_{\mu}^2 \kappa_A^2 }{4 m_W^2  t  } (\bar{v}(p_2)   \slashed{k}_2  u(p_1) + 
\bar{v}(p_2)  \slashed{k}_2  \gamma_5 u(p_1)  + m_{\mu} \bar{v}(p_2)    u(p_1) +   
m_{\mu} \bar{v}(p_2) \gamma_5   u(p_1)  )  \nonumber  \\
M_{0}^{h_i} &=&\bar{v}(p_2) u(p_1)\frac{g_{h_iH^+H^-}}{s-m_{h_i}^2+i m_{h_i} \Gamma_{h_i}} \frac{g m_{\mu} \chi_{h_i}^l}{2 m_W}
\end{eqnarray}
where we have used the  following couplings:
$Z^\mu\mu^+\mu^- = i\,\gamma^\mu\big(g_V-g_A\gamma^5\big)$ 
with $g_V= g (1-4\,s_w^2)/(4\,c_w)$ and $g_A=g/(4\,c_w)$. The $Z$
coupling to a pair of charged Higgs is: 
$g_H=- g (c_w^2-s_w^2)/(2\,c_w)$.\\

We use the following notation:
\begin{eqnarray}\nonumber
Y_V & = &  - \frac{g^2 m_\mu^2 \kappa_A^2 }{4 M_W^2} 
\nonumber\\
a_h &= & \frac{ g_{h H^+ H^-}
 g m_\mu \kappa_{h}^\ell}{2  M_W } \ \ \ , \ \ \
a_H  =  \frac{g_{H H^+ H^-}
g m_\mu \kappa_{H}^\ell}{2 M_W}\nonumber\\
a_V & = & - 2 \frac{e^2}{s} +  \frac{2 g_H g_V}{s-M_Z^2+ i M_Z \Gamma_Z} +
\frac{Y_V}{t} \nonumber  \\
a_A & = &- \frac{2  g_H g_A}{s-M_Z^2+ i M_Z \Gamma_Z} + \frac{Y_V}{t}  \nonumber \\
a_S & = &  \frac{a_h}{s-M_h^2+i M_h \Gamma_h} + \frac{a_H}{s-M_H^2+ i M_H \Gamma_H} +
\frac{m_\mu Y_V }{t} \nonumber \\
a_{SA} & = & \frac{m_\mu Y_V }{t}
\end{eqnarray}
Where we have introduced the total width for the $Z$ boson, $h$ and $H$.
The total widths for $h$ and $H$ 
are computed at the leading order. The inclusion of the total width
is necessary for the case 
where the center-of-mass energy becomes close to the mass of the
neutral-Higgs state: $\sqrt{s}\approx m_H$.
The square of the amplitude is given by:
\begin{eqnarray}
|M|^2 &=& \Big[ \big(|a_V|^2 + |a_A|^2 \big) \frac{s^2}{2}\beta_H^2 \sin^2\theta  + 2\,\big(|a_S|^2 -|a_{SA}|^2\big)\,s \Big] \;,
\end{eqnarray}
where we neglect the muon-mass term in $M_{0}^{\nu}$ amplitude.
The differential cross section is given by:
\begin{eqnarray}
\frac{d\sigma}{d\Omega}=\frac{\beta_H}{64 \pi^2 s}\frac{1}{4} \;.
|M|^2 \label{ref0}
\end{eqnarray}
The factor $1/4$ is due to initial state spin average. It is clear
from above that the amplitude squared of the $s$-channel neutral-Higgs
exchange does not depend on the scattering angle,
therefore such contributions will have a flat angular distribution.

Note that in the case of $e^+e-$ collider, the s-channel with $h$ and $H$ exchange and the t-channel neutrino exchange are neglected because being proportional to the electron mass. The total cross section for $e^+e^- \to H^+ H^-$ 
is given by \cite{Arhrib:1998gr}:
\begin{equation}
\label{eq:tot_cs_ee_hphm}
\sigma_{\rm tot}^{e^+e^-}=\frac{e^4 \pi\alpha^2\beta_H^3}{3s}\bigg( 1 + g_H^2 \frac{g_V^2+g_A^2}{(1-m_Z^2/s)^2}-\frac{2g_Hg_V}{1-m_Z^2/s} \bigg)
\end{equation}

\subsection{$\mu^+ \mu^- \to H^{\pm}W^{\mp}$ production}
\label{section3subsec2}
\begin{figure}[!h]
\centering
\includegraphics[width=0.75\textwidth]{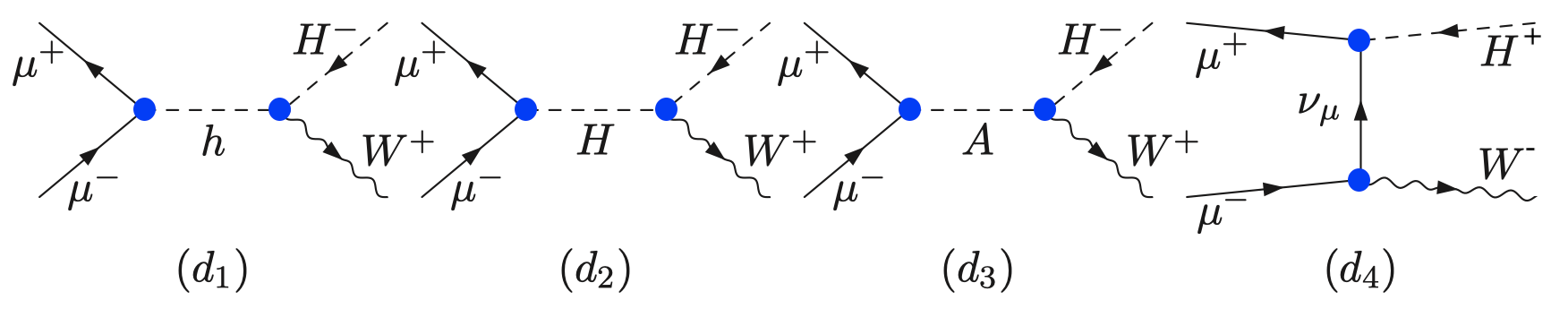}%
\caption{Tree-level Feynman diagrams for $\mu^+ \mu^- \to H^{\pm}W^{\mp}$ at muon collider in the 2HDM}
\label{diag:2mu2HpWm}
\end{figure}

$\mu^+ \mu^- \to H^{\pm}W^{\mp}$ is another process that could be
important for the muon collider. 
Such a process may proceed via an $s$-channel mediated by $h,H$ or $A$ in
Fig.\ref{diag:2mu2HpWm}-$(d_{1,2,3})$ and by neutrino-exchange 
$t$-channel diagram in Fig.\ref{diag:2mu2HpWm}-$(d_4)$.
This process may offer the following possible benefits, 
 (over the standard pair production discussed previously).

\begin{itemize}
\item The process $\mu^+\mu^-\to H^{\pm}W^{\mp}$ can provide information on the underlying Yukawa textures, since all the diagrams presented in Fig.\ref{diag:2mu2HpWm}  contain the muon coupling,
in contrast to $\mu^+\mu^-\to H^+H^-$, which has an almost
  model-independent rate because it is mostly dominated by
  $\mu^+\mu^-\to \gamma^*,Z^*\to H^+H^-$,
  which depend only on the charged Higgs mass and gauge couplings.
\item Compared to charged-Higgs pair production, the production
  of a single $H^{\pm}$ is less constrained by the phase space, which
  further enables for a larger range of kinematics
  at a given center-of0mass energy, with on-shell production possible
  up to $\sim \sqrt s-M_W$.
\item The process may also experience resonance enhancement from the
  $s$-channel heavy-Higgs exchanges of $H$ and $A$ along the
  following patheways: $\mu^+\mu^-\to H^*, A^*\to  H^{\pm}W^{\mp}$.
  Furthermore, this could be significant given that the coupling
  $W^\pm H^\mp A$ is a gauge coupling without any mixing suppression
  and $W^\pm H^\mp H$ is proportional to $\sin(\beta -\alpha)$,
  which is driven by the LHC data to its maximum value
  $\sin(\beta-\alpha)\approx 1$  \cite{ATLAS:2022vkf, CMS:2022dwd}.
\item
  The contribution from the $t$-channel may be substantial regardless of
  whether $\sqrt s\approx M_{H,A}$ or not.
\end{itemize}

The kinematic for $\mu^+\mu^- \to H^\pm W^\mp$ is fixed as follow. 
The momenta of the incoming $\mu^+$ and $\mu^-$, the outgoing charged Higgs
boson $H^{\pm}$ and the gauge boson $W^{\mp}$  are denoted by $p_{1,2}$ and
$k_{1,2}$, respectively.  Neglecting the muon mass $m_{\mu}$,
the momenta in the $\mu^+ \mu^-$ center-of-mass  system are given by:
\begin{eqnarray}
& & p_{1,2}=\frac{\sqrt{s}}{2} (1,0,0,\pm 1) \nonumber \\
& & k_{1, 2}=\frac{\sqrt{s}}{2} (1\pm \frac{m_W^2 -m_{H^\pm}^2}{s},
\pm \frac{1}{s}\lambda^{\frac{1}{2}}(s,m_W^2,m_{H^\pm}^2) \sin\theta,0,\pm 
\frac{1}{s}\lambda^{\frac{1}{2}}(s,m_W^2,m_{H^\pm}^2)
\cos\theta), \nonumber
\end{eqnarray}
\noindent
Here $\lambda(x,y,z)=x^2+y^2+z^2-2xy-2xz-2yz$ is the usual two body phase space function and
$\theta$ is the scattering angle between $\mu^+$ and $H^+$. The $s,\, t$ and
$u$ can be written as:
\begin{eqnarray}
& & s =  (p_{1}+p_{2})^2 = (k_1+k_{2})^2  \nonumber\\
& & t = (p_{1}-k_1)^2 = (p_{2}-k_{2})^2 =
\frac{1}{2}(m_W^2  + m_{H^\pm}^2) -
\frac{s}{2}
+\frac{1}{2} \lambda^{\frac{1}{2}}(s,m_W^2,m_{H^\pm}^2)
\cos\theta  \nonumber\\
& & u = (p_{1}-k_{2})^2 = (p_{2}-k_1)^2 =
\frac{1}{2} (m_W^2 + m_{H^\pm}^2) -
\frac{s}{2}-
\frac{1}{2} \lambda^{\frac{1}{2}}(s,m_W^2,m_{H^\pm}^2) \cos\theta \\
& & s+t+u = m_W^2 + m_{H^\pm}^2  \nonumber
\end{eqnarray}

Similar to the first process, the $s$-channel  $M_{0}^{h,H}$, $M_{0}^{A}$  
and $t$-channel $M_{0}^{\nu}$ amplitudes are respectively given by:
\begin{eqnarray}
%  \slashed{k}_2
M_{0}^{h} &=&    \frac{ g^2 m_{\mu} }{4 m_W } \frac{\cos(\beta-\alpha)\kappa_h^l }{s-m_h^2 + i m_h \Gamma_h}  
 \bar{v}(p_2)  u(p_1) (2 k_1 +k_2)^\mu \epsilon_{\mu}(k_2) \nonumber  \\
 M_{0}^{H} &=&    \frac{ g^2 m_{\mu} }{4 m_W } \frac{-\sin(\beta-\alpha) \kappa_H^l}{s-m_H^2 + i m_H \Gamma_H}  
 \bar{v}(p_2)  u(p_1) (2 k_1 +k_2)^\mu \epsilon_{\mu}(k_2)  \nonumber  \\
 M_{0}^{A} &=&    \frac{ g^2 m_{\mu} }{4 m_W } \frac{- \kappa_A^l}{s-m_A^2 + i m_A \Gamma_A}  
 \bar{v}(p_2)  u(p_1) (2 k_1 +k_2)^\mu \epsilon_{\mu}(k_2)  \nonumber   \\
 M_{0}^{\nu} &=&    \frac{g^2 m_{\mu} }{2 m_W} \frac{\kappa_A^l}{t}  
 \bar{v}(p_2) \gamma^\mu \frac{1-\gamma_5}{2}  ( \slashed{k}_2- \slashed{p}_2 ) u(p_1)  \epsilon_{\mu}(k_2)  
\end{eqnarray}
Taking into account the spin average of the initial state and
polarization sum of the $W$ gauge boson,  
the square of the amplitude is given by:
\begin{eqnarray}
|{\cal M}|^2 & = &  \frac{s g^4 m_{\mu}^2 }{32m_W^4}
\Big[ \big(|a_V|^2+|a_A|^2\big) \lambda( s,m_{H^\pm}^2,m_W^2) + 2 a_t^2 \big(2 M_W^2 p_T^2 + t^2 \big)  \nonumber\\ 
&+&2 a_t \big(m_{H^\pm}^2 m_W^2-s p_T^2-t^2\big) \Re(a_V  -a_A) \Big]
\end{eqnarray}
where $s p^2_T=tu - m_W^2 m_{H^\pm}^2 = \lambda(s,M_{H^{\pm}}^2,M_W^2) \sin^2\theta/4$, while the couplings $a_V$ and $a_A$ are given by:
\begin{eqnarray}
a_V &=& \left( { \cos(\beta-\alpha) \kappa_{h}^{\ell} \over s-m_h^2+
im_h\Gamma_h}
-{ \sin(\beta-\alpha) \kappa_{H}^{\ell} \over s-m_H^2+im_H\Gamma_H}\right)\nonumber \\
a_A &=& {\kappa_{A}^{\ell}\over s-m_A^2+im_A\Gamma_A}  \qquad ; \qquad
a_t = {\kappa_{A}^{\ell}\over t}  \;.
\end{eqnarray}
The differential cross-section for
$\sigma(\mu^+\mu^-\to H^{\pm}W^{\mp})$ may be written as follows:
\begin{equation}
{d\sigma\over d\Omega} = {\lambda^{1\over 2}(s,m_{H^{\pm}}^2,m_W^2)\over
64\pi^2s^2} |{\cal M}|^2 \label{ref1}
\end{equation}
It is clear from the above amplitude squared of the $s$-channel neutral-Higgs
exchange diagrams does not have any 
scattering angle dependence. Therefore, such $s$-channel Higgs-exchange
diagrams will have a flat angular distribution.

\begin{figure}[!ht]
 \centering
 \includegraphics[scale=0.37]{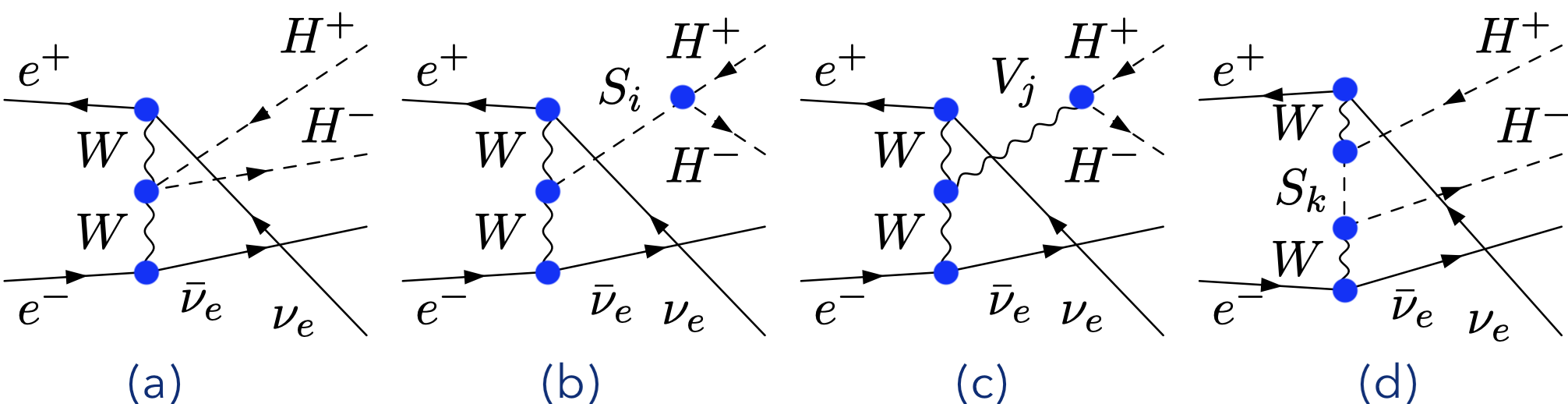}
 \caption{Feynman diagrams involving di-charged Higgs production from VBF at the ILC $e^+e^- \to \nu \bar{\nu} H^+ H^-$. 
 $S_i$, $V_j$ and $S_k$ refer to $(h, H)$, $(\gamma,\,Z)$ and $(h, H, A)$ respectively.}
 \label{fig:diag-VBF-HpHm}
 \end{figure}

\noindent
\underline{Vector Boson Fusion (VBF)}\\ 
Before ending this section, and in addition to the above, it's worth mentioning that vector boson fusion (VBF) processes: $e^+e^- \to \nu \bar{\nu} H^+ H^-$
 hold significant potential for producing charged Higgs bosons at electron-positron colliders, whether at the ILC with a center-of-mass energy of 500 GeV$\sim$1 TeV or at the CLIC with energies of 3 TeV, 6 TeV, and 10 TeV. While our initial analysis focused on the muon collider, we recognize that VBF processes could contribute significantly in certain scenarios. Specifically, the production of charged Higgs bosons alongside neutrinos via VBF in electron-positron collisions could occur at significant rates, depending on factors like collider energy and luminosity. The corresponding Feynman diagrams are shown in the Figure.\ref{fig:diag-VBF-HpHm}

\section{Numerical results}
\label{Section4}
\begin{figure}[htb!]
\centering
% for HpHm production
\includegraphics[width=0.325\textwidth]{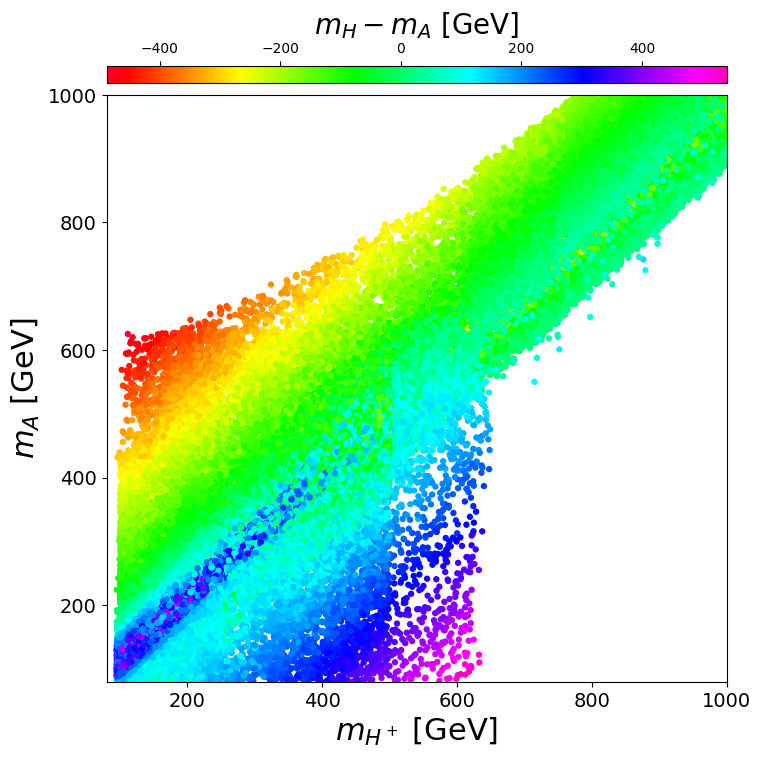}
\includegraphics[width=0.325\textwidth]{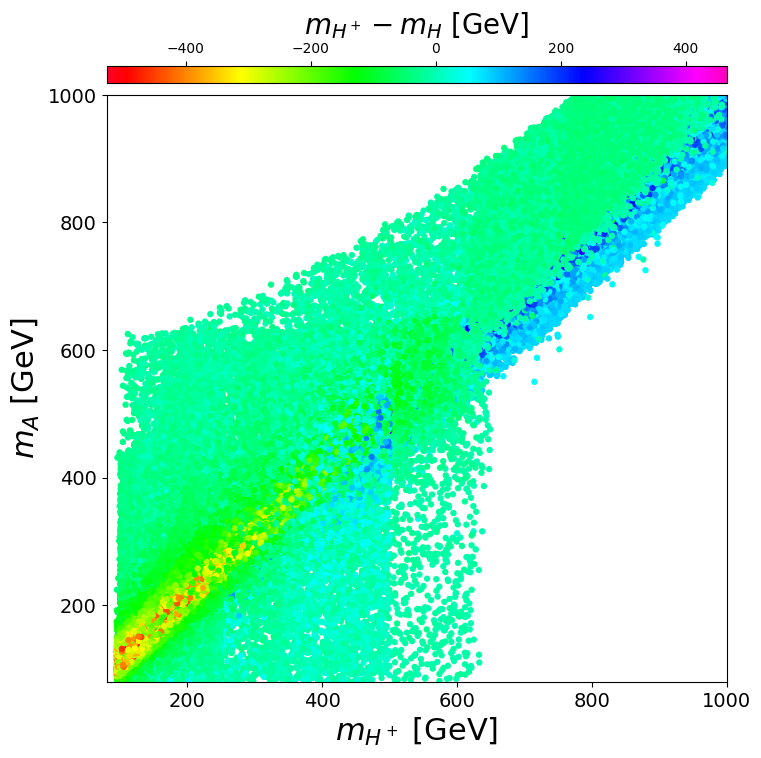}
\includegraphics[width=0.325\textwidth]{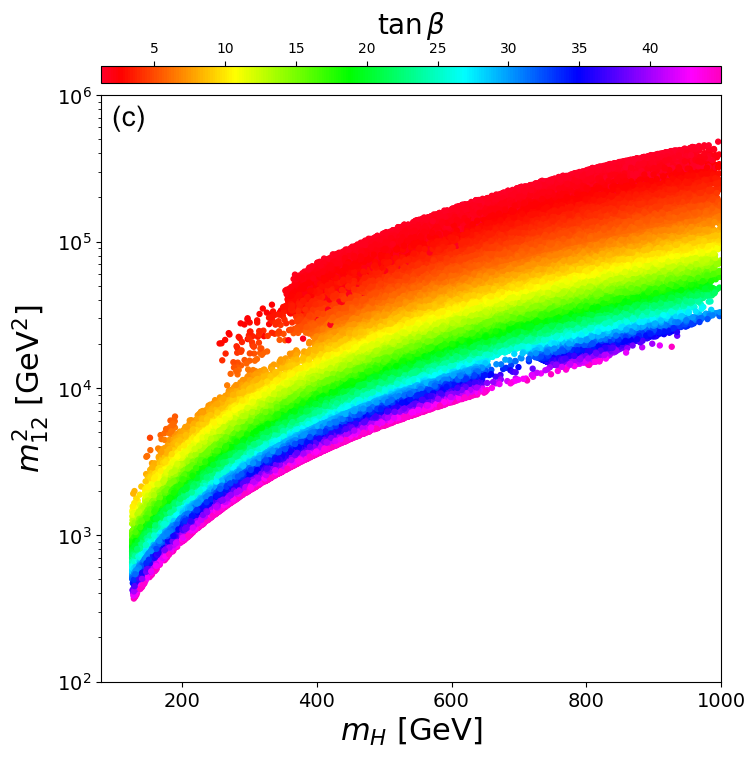}
\caption{Correlations between $m_A$ and $m_{H^\pm}$ (left and middle) and between $m_{12}^2$ and $m_H$ (right) after imposing the theoretical and experimental constraints. The color code indicates the splitting $m_H - m_A$, $m_{H^+} - m_H$ and $\tan\beta$, respectively.}
\label{correlation}
\end{figure}

We perform random scans over the parameter space of the 2HDM
within the following ranges:
\begin{eqnarray}
\centering
&&m_{h} = 125.09\ \text{GeV}, \ \ m_{H} \in [130,~1000]\ \text{GeV},\hspace{0.3cm} \sin(\beta-\alpha)\in [0.97,1], \ \  \nonumber \\ && m_{A, H^\pm}\in [80,~1000]\ \text{GeV}, 
  \ \  \tan\beta\in [0.5,~45],\  \hspace{0.3cm}m_{12}^2 \in [0,1000^2]\;,
\label{parm}
\end{eqnarray}
where we have assumed that the lightest Higgs state $h$ is
the observed SM-like Higgs boson at the LHC 
\cite{ATLAS:2012yve,CMS:2012qbp}  and set $m_h=125$ GeV. 
After scrutinizing the parameter space of the model with the
theoretical and experimental constraints described above,
the resulting parameter space points  will be passed to
FormCalc \cite{Hahn:2001rv,Hahn:1998yk,Kublbeck:1990xc} to compute
the corresponding cross section of each process at the muon collider.  

\begin{figure}[htb!]
\centering
\includegraphics[width=0.325\textwidth]{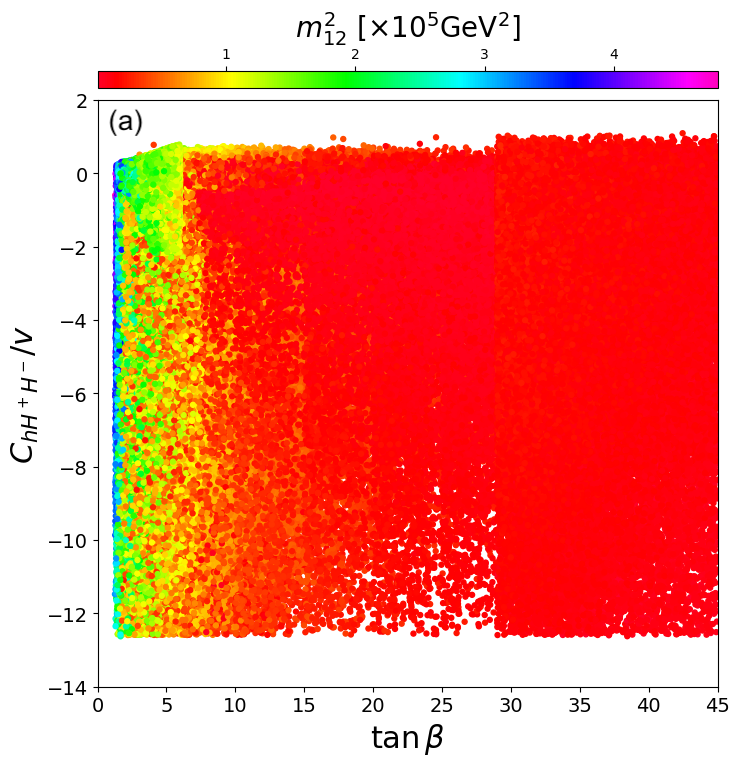}
\includegraphics[width=0.325\textwidth]{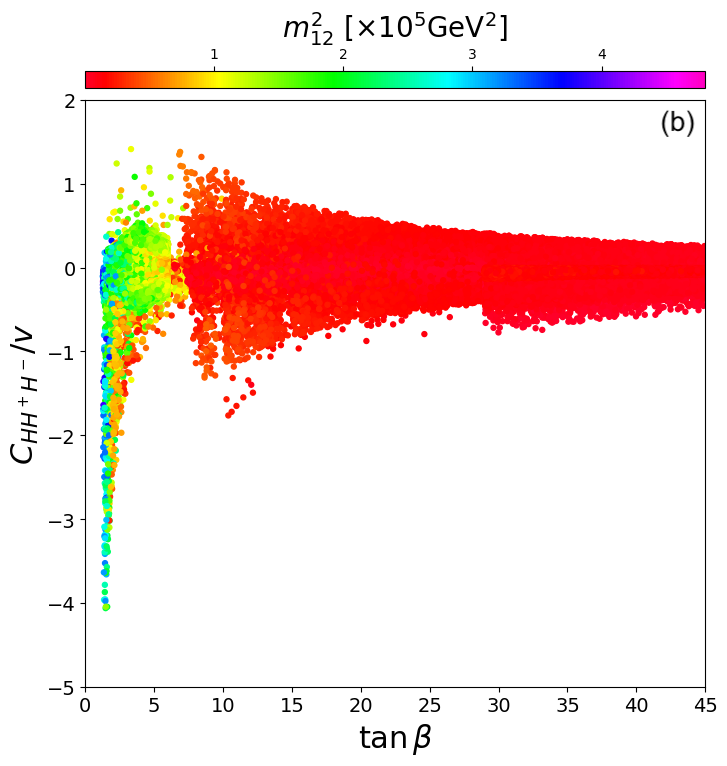}
\includegraphics[width=0.325\textwidth]{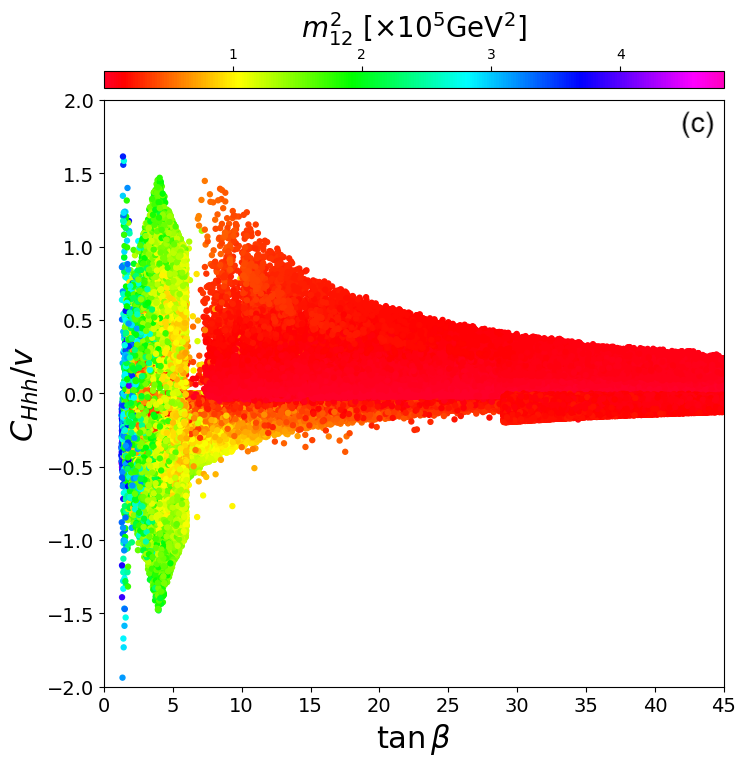}
\caption{Pure scalar couplings normalized to the SM vev $v$
  as a function of $\tan\beta$: (a) $g_{hH^+H^-}/v$, (b) $g_{HH^+H^-}/v$,  and
  (c) $g_{Hhh}/v$ with $m_{12}^2$ represented by the color code. }
\label{coupling}
\end{figure}

In the plane $[m_A,\,m_{H^\pm}]$, we illustrate the mass splitting between 
$m_{H}$ and $m_A$ in Fig.\ref{correlation}(a) and between $m_{H^\pm}$ and $m_H$
in Fig.\ref{correlation}(b), while Fig.\ref{correlation}(c) exhibits the correlation between $m_H$ and $m_{12}^2$.
It is clear that for $m_{H^\pm}\leq 600$ GeV, the splitting between 
$A$ and $H$ could be quite large while for $m_{H^\pm}\geq 600$ GeV,
the splitting between the charged Higgs and neutral heavy states
$H$ and $A$ becomes rather small.
In Fig.\ref{correlation}(c) we show the correlation between
$m_H$ and $m_{12}^2$, as we know the mass of the heavy states
$H$, $A$ and $H^\pm$ are driven by $m_{12}^2$ \cite{Gunion:2002zf}:
\begin{eqnarray}
m_{H,A,H^\pm}^2 \approx \frac{m_{12}^2}{s_{\beta} c_{\beta}} + \lambda_i v^2 + {\cal{O}}(v^4/m_{12}^2) \;.
\label{expanssio} 
\end{eqnarray}
In order to have $m_H$ of the order 1 TeV,
a large $m_{12}^2\geq 3\times 10^4$ GeV$^2$ is needed.

In Fig.\ref{coupling}, we illustrate the pure scalar couplings
normalized to the vev as functions of $\tan\beta$.
In the panels (a), (b) and (c), one can see the allowed size for
$g_{hH^+H^-}/v$,  $g_{HH^+H^-}/v$ and  $g_{Hhh}/v$, respectively.
To understand this behavior, we give the couplings $g_{hH^+H^-}$, $g_{HH^+H^-}$
and $g_{Hhh}$ of Eqs.~(\ref{thdm:h1hphm}) in the alignment limit
$\sin(\beta-\alpha)\approx 1$ ($\beta\approx \alpha+\pi/2$) as 
\begin{eqnarray}
g_{h H^+H^-}& = & -\frac{1}{v}\bigg[  2m_{H^\pm}^2+m_h^2  -2 \frac{m_{12}^2}{s_\beta c_\beta}  \bigg] \nonumber\\
g_{H H^+H^-}& = & -\frac{1}{v}\bigg[ -\Big(m_H^2 - \frac{m_{12}^2}{s_\beta c_\beta} \Big) \frac{\cos (2 \beta)}{s_\beta c_\beta } \bigg] \nonumber\\
g_{Hhh}&=&  -\frac{c_{\beta-\alpha}}{v }\bigg[  - (2 m_h^2 + m_H^2)   + 4 \frac{m_{12}^2}{s_\beta c_\beta} \bigg] 
\end{eqnarray}
Further in the large $\tan\beta$ limit, we have: $1/(s_\beta c_\beta)\approx \tan\beta +  {\cal{O}} (\frac{1}{\tan\beta})$, $\cos(2\beta)/(s_\beta c_\beta)\approx -\tan\beta +  {\cal{O}} (\frac{1}{\tan\beta})$ and  $\cos(2\beta)/(s_\beta^2 c_\beta^2)\approx -\tan^2\beta +  {\cal{O}} (\frac{1}{\tan^2\beta})$, therefore:
\begin{eqnarray}
g_{h H^+H^-}& \approx & -\frac{1}{v}\bigg[  2m_{H^\pm}^2+m_h^2  -2 m_{12}^2 ( \tan\beta +{\cal{O}} (\frac{1}{\tan\beta}) )   \bigg] \nonumber \\
g_{H H^+H^-}& \approx  & -\frac{1}{v}\bigg[ - m_H^2 ( -\tan\beta + {\cal{O}} (\frac{1}{\tan\beta}) ) + m_{12}^2 ( -\tan^2\beta + {\cal{O}} (\frac{1}{\tan^2\beta}) ) \bigg] \\
g_{Hhh}&\approx& -\frac{c_{\beta-\alpha}}{v }\bigg[  - (2 m_h^2 + m_H^2)   +  4 m_{12}^2 \tan\beta  + {\cal{O}} (\frac{1}{\tan^2\beta})\bigg] \;.
\end{eqnarray}
It is evident that both $g_{hH^+H^-}$ and $g_{Hhh}$ have partial linear
dependence on $\tan\beta$, while $g_{HH^+H^-}$ has both partially linear
and quadratic dependence on $\tan\beta$.  However, this $\tan\beta$
dependence will eventually cancel out, because both $m_H$ and $m_{H^\pm}$
are given by $ \frac{m_{12}^2}{s_{\beta} c_{\beta}} \approx m_{12}^2
\tan\beta$ in the large $m_{12}^2$ limit. This cancellation is more
pronounced for $g_{HH^+H^-}$ and $g_{Hhh}$ couplings, where it can be observed
that, for large $\tan\beta$, the $g_{HH^+H^-}$ and $g_{Hhh}$ couplings become
small. 
%===================
\begin{figure}[htb]
\centering
\includegraphics[width=0.35\textwidth]{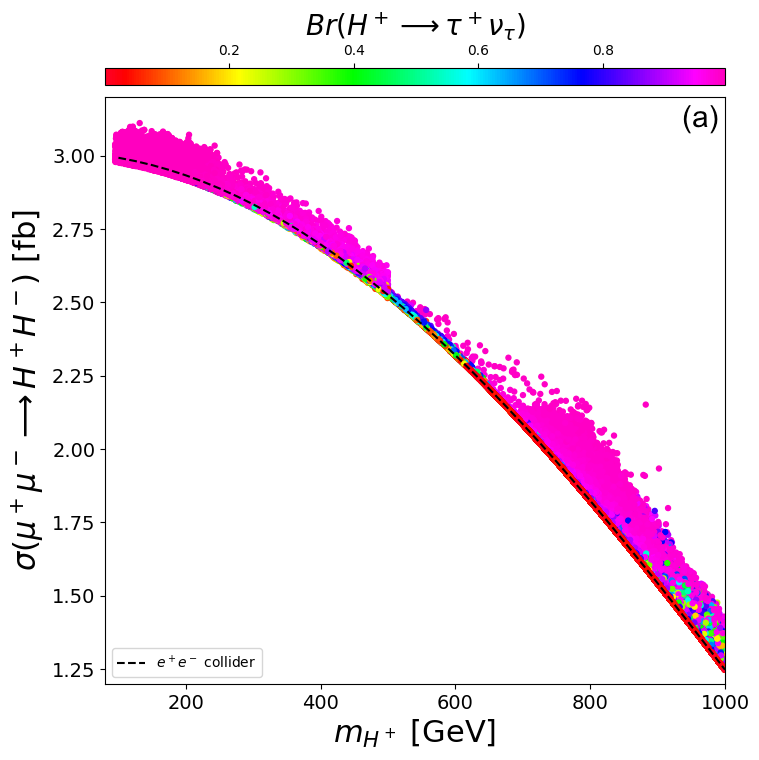}
\includegraphics[width=0.35\textwidth]{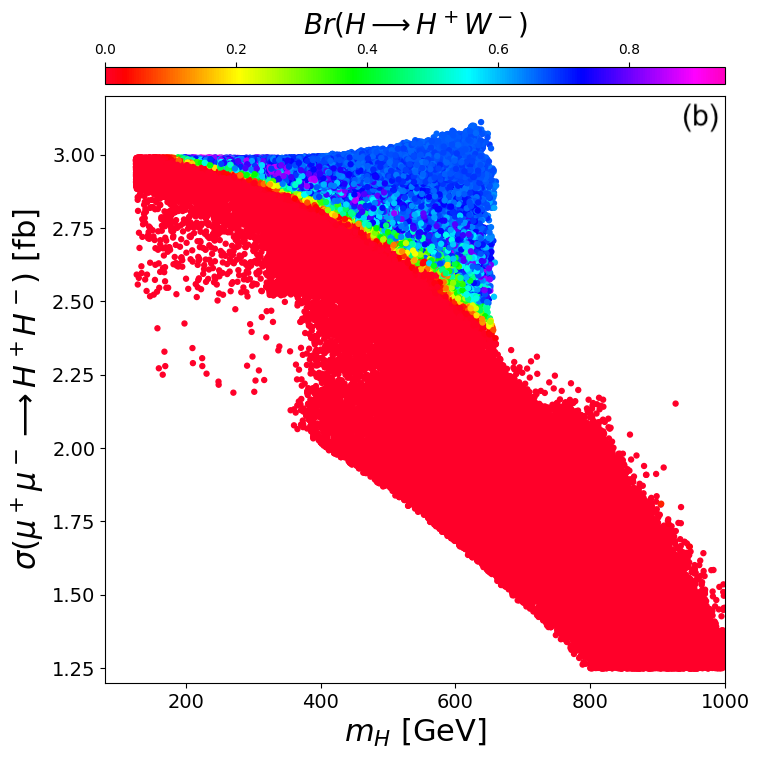}\\
\includegraphics[width=0.35\textwidth]{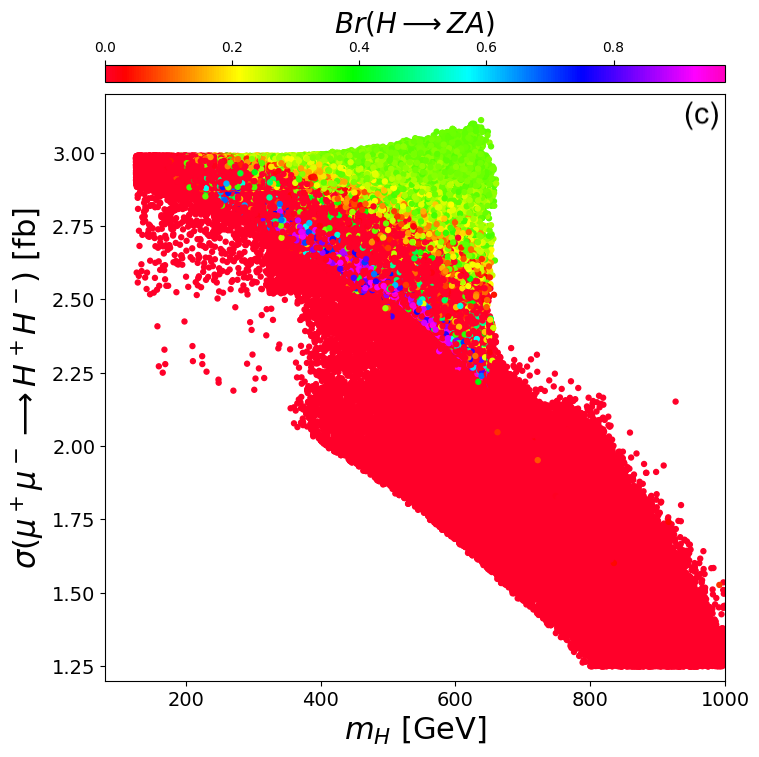}
\includegraphics[width=0.35\textwidth]{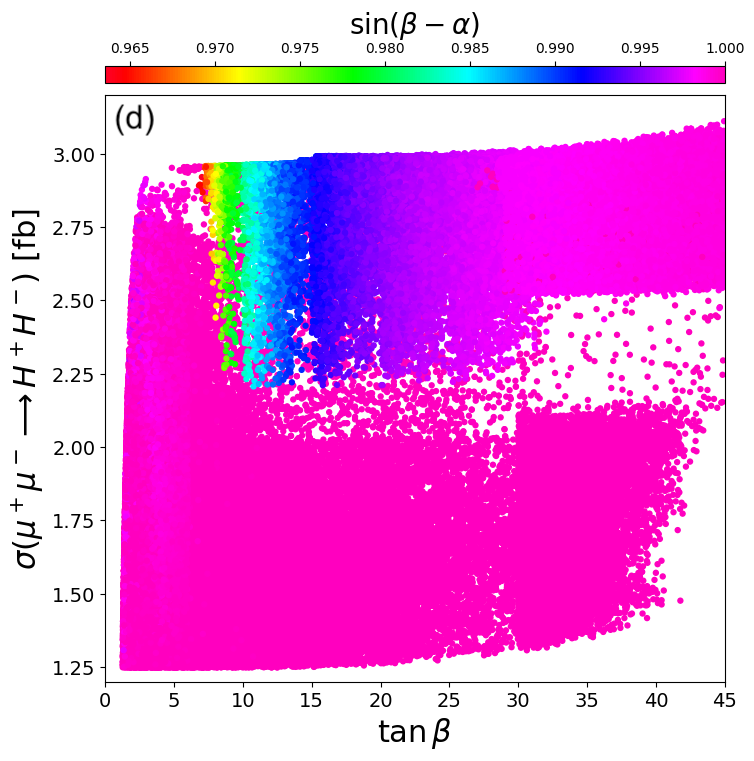}
\caption{Production cross sections for $\mu^+\mu^- \to H^+ H^-$ in Type X as a function of $m_{H^\pm}$, $m_H$ and $\tan \beta$, at $\sqrt{s}=3$ TeV. The color legend presents respectively the following observables: $\sin(\beta - \alpha)$, $Br(H^+ \to \tau^+\bar{\nu_\tau})$, $Br(H \to H^\pm W^\mp)$ and $Br(H \to ZA)$. All the regions are consistent with theoretical and experimental constraints.}
\label{cs_results_hh}
\end{figure}

\begin{itemize}
\item $\mu^+ \mu^- \to H^+H^-$
\end{itemize}

The key results of our investigation for the muon collider with
center-of-mass energy of  3 TeV are illustrated in Fig.\ref{cs_results_hh} for charged-Higgs pair production. For such process, we first point out that because the dominant contribution comes from the $s$-channel $\mu^+ \mu^- \to \gamma, Z \to H^+H^-$ diagrams (Fig.\ref{diag:2mu2HpHm}$(d_{3,4})$), the cross section is exactly the same as one can obtain via $e^+e^- \to H^+ H^-$ at $e^+e^-$ colliders at the same $\sqrt{s}$.
In additional, we have the $s$-channel neutral-Higgs exchange
$\mu^+ \mu^- \to h, H \to H^+H^-$ (Fig. \ref{diag:2mu2HpHm}$(d_{1,2})$),
which is in general smaller than $\mu^+ \mu^- \to \gamma, Z \to H^+H^-$. 
%but also make a positive and constructive contribution to the total cross section.
%
The $t$-channel contribution is
rather suppressed by $m_{\mu}^4$. We stress here that the interference
between $\mu^+ \mu^- \to \gamma, Z \to H^+H^-$ and $\mu^+ \mu^- \to
h,H \to H^+H^-$ is constructive, resulting in a cross section that is
all time larger than the corresponding $\sigma(\mu^+ \mu^- \to \gamma,
Z \to H^+H^-)$=$\sigma(e^+e^- \to H^+ H^-)$.

%The upper panels in Fig.\ref{cs_results_hh} show the correlations
%between the production cross section $\sigma(\mu^+ \mu^- \to H^\pm
%H^\mp)$ and $m_{H^\pm}$, $m_{H}$, and $\tan\beta$ for
%$\sqrt{s}=500$ GeV and while those for $\sqrt{s}=1$ TeV are shown in 
%lower panels.

The panels in Fig.\ref{cs_results_hh} show the correlations
between the production cross section $\sigma(\mu^+ \mu^- \to H^\pm
H^\mp)$ and $m_{H}$, $m_{H^\pm}$ and $\tan\beta$ for
$\sqrt{s}=3$ TeV.

The cross section is slightly improved overall in the permitted dataset. It can reach a maximum value - up to $\sim3.2$ fb for low $m_{H^\pm}$ however large $\tan\beta$. Keep in mind that the 2HDM type II and type-X (or type III) are both affected
by the enhancement for large $\tan\beta$ (see Table~\ref{coupIII}  for the
couplings).  We also see that, as supported by the LHC data, the
maximal cross section is found for $\sin(\beta-\alpha) \approx
1$. From the plots in Fig.\ref{cs_results_hh}-(b) and (c), one can clearly see the
large rise in the total cross section,
which corresponds to the
large $\tan\beta$ from the $s$-channel contribution
$\mu^+ \mu^- \to H^* \to H^+ H^-$ when $H\mu^+ \mu^-$ receives significant $\tan\beta$ amplification.

Furthermore, it can be observed from Fig.\ref{cs_results_hh}(a), that most of the time, there is constructive interference between the ($\gamma$,\,$Z$) s-channel and ($h$,\,$H$) s-channel contributions, since the cross section for $\mu^+ \mu^- \to H^+ H^-$ is greater than $e^+ e^- \to H^+ H^-$. However, such interference could also be destructive, as seen in a small region of the parameter space in Fig.\ref{cs_results_hh}(a) below the dashed line, where $\mu^+ \mu^- \to H^+ H^-$ is smaller compared to $e^+ e^- \to H^+ H^-$.
  
%As can be observed, 
Additionally, the maximum enhancement is achieved for the
neutral-Higgs mass $M_{H}$ in the region of 
$[500, 650]$ GeV and the highest allowable $\tan\beta \approx 45$.
Here, we emphasize that the theoretical and experimental constraints
still permit $\tan\beta\geq 45$ for 2HDM type X,
suggesting that the cross section may be a little larger than what we show.
We only scan up to $\tan\beta \leq 45$ in our study
due to time constraints.
The cross section $e^+e^- \to H^+ H^-$ starts to become suppressed in the
range $m_{H^\pm} \in [200, 250]$ GeV due to phase space suppression, see 
Fig.\ref{cs_results_hh}(a) (dash line). As a result, 
the muon collider can be used in addition to $e^+e^-$ machines
to study the region of the parameter space that the linear collider is unable to access.

The CP even Higgs can decay into $b\bar{b}$, $\tau^+\tau^-$, $WW$,
$ZZ$, $t\bar{t}$, $ZA$, $hh$, $W^\pm H^\mp$, and $H^+H^-$. In the 2HDM
Type X, $H\to \tau^+ \tau^-$ would be the dominant decay mode
 at large $\tan\beta$.  $H\to WW, ZZ$ are suppressed since both
are proportional to $\cos(\beta-\alpha)\approx 0$, but nevertheless
they could reach a few percent branching fraction in some cases.
After crossing the $t\bar{t}$ threshold, there is a strong competition
between $hh$, $ZA$, $W^\pm H^\mp$, and $H^+H^-$.  The decay channel
$H\to H^+H^-$, which is open only for $m_H> 2 m_{H^\pm}$, is rather small
compared to $H\to W^\pm H^\mp $ and $H\to Z A$, which have more phase
space and the coupling $HW^\pm H^\mp, HZA \propto \sin(\beta-\alpha)$
is maximal and this makes the $Br(H\to W^\pm H^\mp)$ and $Br(H\to ZA)$
rather substantial as can be seen from Figs.\ref{cs_results_hh}(b) and (c).  Therefore, since $Br(H\to H^+H^-)$ is very small,
the cross section $\sigma(\mu^+ \mu^- \to H^* \to H^+H^-)$ is not
strongly enhanced because
$\sigma(\mu^+ \mu^- \to H^* \to H^+H^-)\approx \sigma(\mu^+ \mu^- \to
H) \times Br(H\to H^+ H^-)$. On the other hand, one would expect that the cross section  
$\sigma(\mu^+ \mu^- \to W^\pm H^\mp)$ would receive
significant enhancement.

In the Figure~\ref{cs_results_hh}, one can also read
branching fractions of $H^\pm$. For $m_{H^+}\equiv86\sim120$ GeV, the
decay mode $H^+ \to \tau \nu_\tau$ dominates. However, as $m_{H^+}$
gets larger, e.g. $m_{H^+} \ge 175$ GeV, a significant competition
arises among the decay modes $H^+\to t\bar{b}$, $H^+\to W^+A$, $H^+\to
W^+h$, and $H^+\to W^+H$.

It is evident that as the center-of-mass energy grows  from $\sqrt{s}=3$
TeV to 10 TeV , the cross section reduces considerably. The cross
section is dominated by the $s$-channel diagrams $\mu^+ \mu^- \to
\gamma, Z \to H^+H^-$ and $\mu^+ \mu^- \to h,A \to H^+H^-$, both of which
behave like $1/s$. 
Note that at these center-of-mass energies, the cross section
$\mu^+ \mu^- \to H^+H^-$ is fully dominated by the $s$-channel $\gamma$
and $Z$ boson exchanges while the $s$-channel Higgs exchange is
much smaller than the other two,
except near the charged Higgs pair production threshold, where both
$s$-channel become comparable.
For the charged-Higgs mass increase  from 250 GeV to 1 TeV and small $\tan\beta$, the cross section decreases from 3 fb to 1.5 fb. The muon coupling to the neutral Higgs bosons $h,H$
can increase the cross section for large $\tan\beta$. 
\begin{itemize}
\item $\mu^+\mu^- \to H^\pm W^\mp$
\end{itemize}

 %===================
\begin{figure}[!h]
\centering
\includegraphics[width=0.325\textwidth]{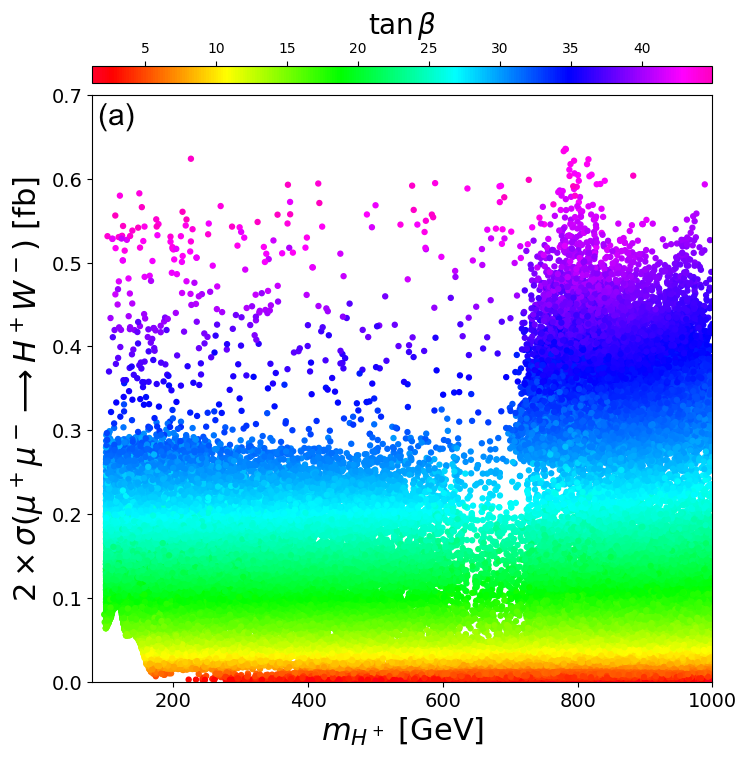}
\includegraphics[width=0.325\textwidth]{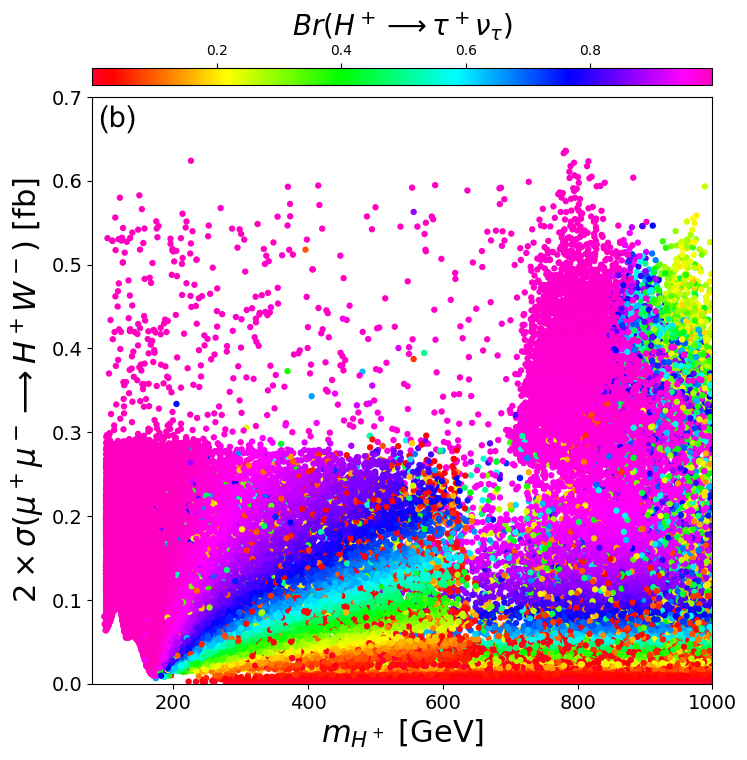}
\includegraphics[width=0.325\textwidth]{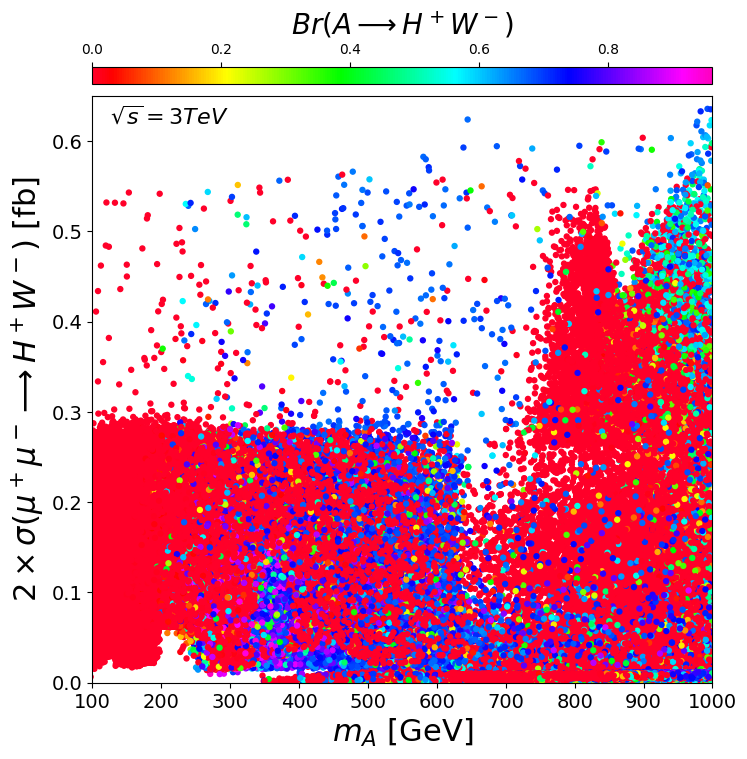}
\caption{Production cross sections for $\mu^+\mu^- \to W^\pm H^\mp$ in Type X as a function of $m_{H^\pm}$ (left and middle panels) and $m_{A}$ (right panel), at $\sqrt{s}=3$ TeV. The color legend presents respectively the following parameter/observables: $\tan\beta$, $Br(H^+ \to \tau^+\bar{\nu})$ and $Br(A \to H^\pm W^{\mp})$. All the regions are consistent with theoretical and experimental constraints.}
\label{cs_results_wh}
\end{figure}

The other part of our investigation for the muon collider at $\sqrt{s}=3$ TeV are illustrated Fig.\ref{cs_results_wh} for the associate production of a charged Higgs with $W$. Unlike $\mu^+\mu^- \to H^\pm H^\mp$, the present process is insensitive to the pure scalar couplings due to the presence of $W$ in the final state. In addition, the $s$-channel diagram with the exchange of $h$ is suppressed by $hW^\pm H^\mp \propto \cos(\beta-\alpha)\approx 0$, while the $s$-channel diagrams with $H$ and $A$ exchanges will benefit from an enhancement since $HW^\pm H^\mp$ is proportional to $\sin(\beta-\alpha)\approx 1$ while $AW^\pm H^\mp$ is a pure gauge coupling
without any mixing factor.

Unlike $\mu^+\mu^- \to H^\pm H^\mp$, the
$t$-channel contribution for $\mu^+\mu^- \to W^\pm H^\mp$ with neutrino
exchange is smaller than the $s$-channel contribution involving the
neutral-Higgs exchanges but is not negligible.
In the case of $\mu^+\mu^- \to H^\pm H^\mp$, the $t$-channel
amplitude is proportional
to $m_{\mu}^2$, while for $\mu^+\mu^- \to W^\pm H^\mp$ the $t$-channel
amplitude has only one $m_{\mu}$ suppression which could overcome
with the large $\tan\beta$ value. Therefore, both the $t$- and
$s$-channel contributions are proportional to $\tan\beta$ in the
large $\tan\beta$ limit, since both $H\mu^+\mu^- \propto
c_{\alpha}/c_{\beta}\approx c_{\beta-\alpha}+\tan\beta
s_{\beta-\alpha}\approx \tan\beta$.
In the large $\tan\beta$ limit,
the amplitudes of all diagrams are proportional to $\tan\beta$,
therefore we would expect enhancement for large $\tan\beta$ 
for both Type II and Type X.

Let's recall that the CP-odd boson $A$ can decay
into $b\bar{b}$, $\tau^+ \tau^-$, $t\bar{t}$, $Zh$, $ZH$, and
$W^\pm H^\pm$. Due to CP invariance, $A$ does not couple to
$WW$ and $ZZ$. Before
the opening of the $t\bar{t}$ threshold, the CP-odd $A$ would decay
dominantly into $\tau^+ \tau^-$ and $b\bar{b}$ with a preference into
$\tau^+ \tau^-$ for large $\tan\beta$,
while for $m_A>2m_t$ there will
be a strong competition between $A\to t\bar{t}$, $A\to \tau^+ \tau^-$,
$A\to ZH$, and $A\to W^\pm H^\pm$ channels since $A\to Zh$ is
proportional to $\cos(\beta-\alpha)\approx 0$ while $A\to t\bar{t}$ is
proportional to $1/\tan\beta$ which is suppressed for large
$\tan\beta$ limit. As the plots indicate, $Br(A\to W^\pm H^\mp)$ is
the dominant one and could exceed 70\%. The possible decays of $H$
have been discussed above.

 \begin{figure}[!h]
 \begin{center}
 \includegraphics[scale=0.265]{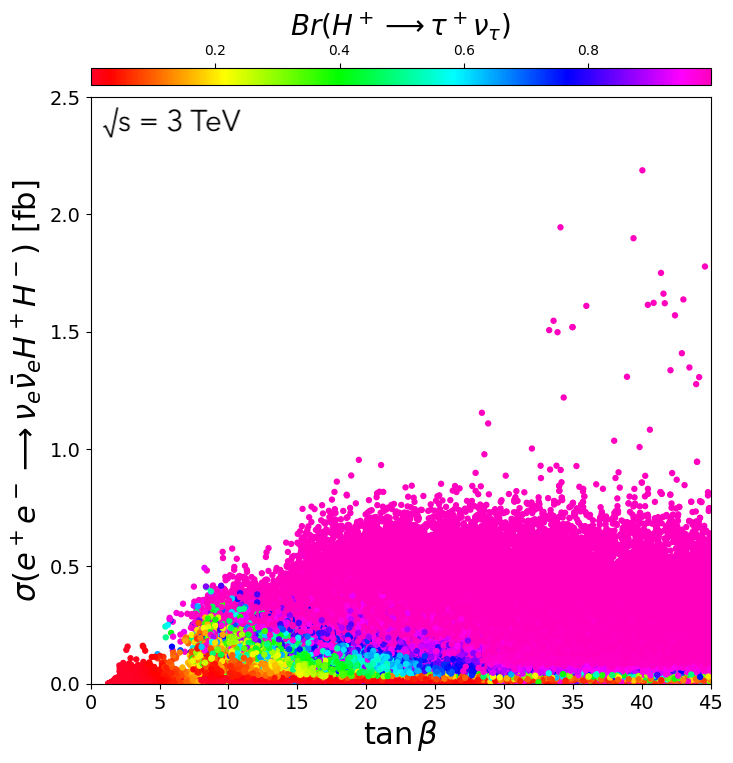}
 \includegraphics[scale=0.265]{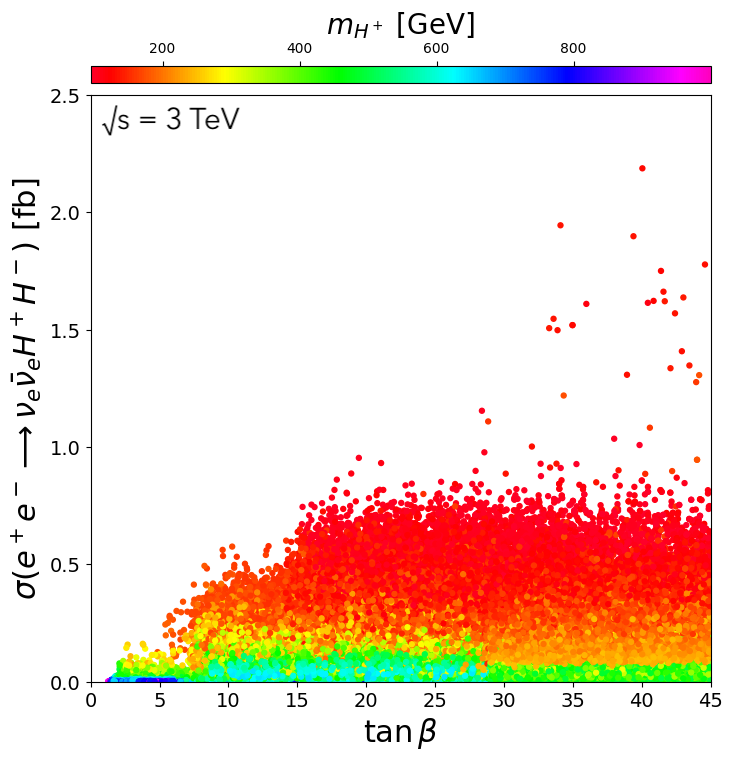}
 \includegraphics[scale=0.265]{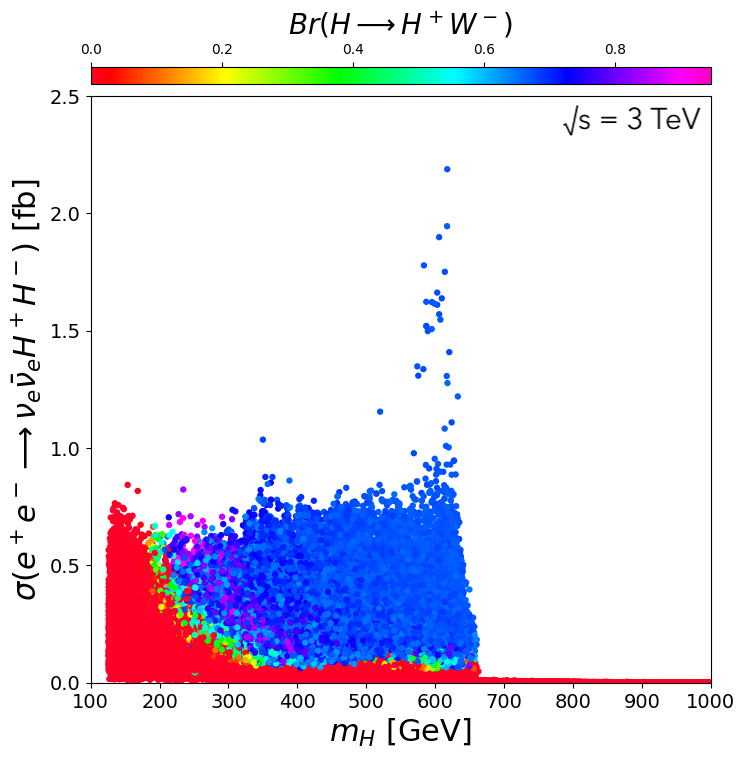}\\
 \includegraphics[scale=0.265]{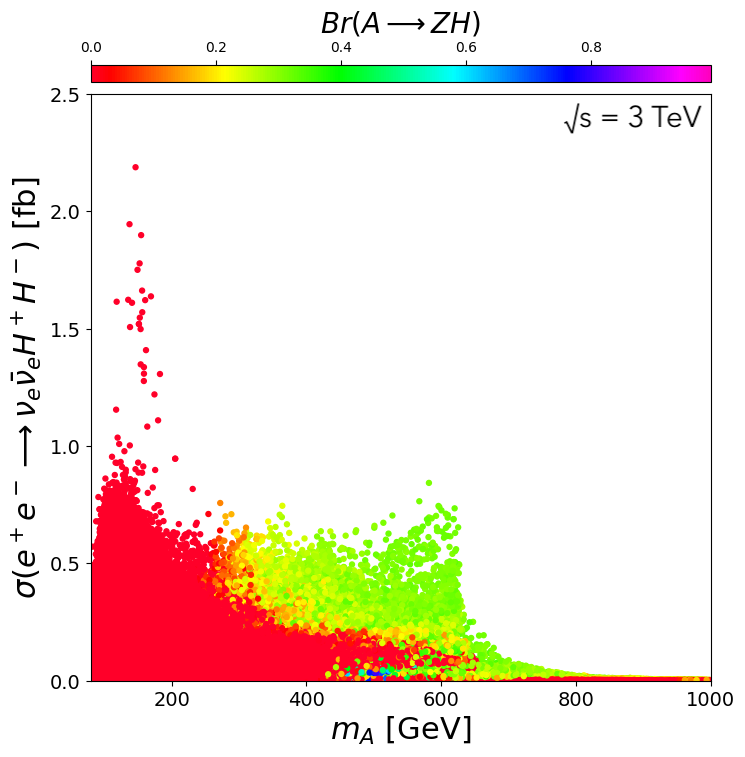}
 \includegraphics[scale=0.265]{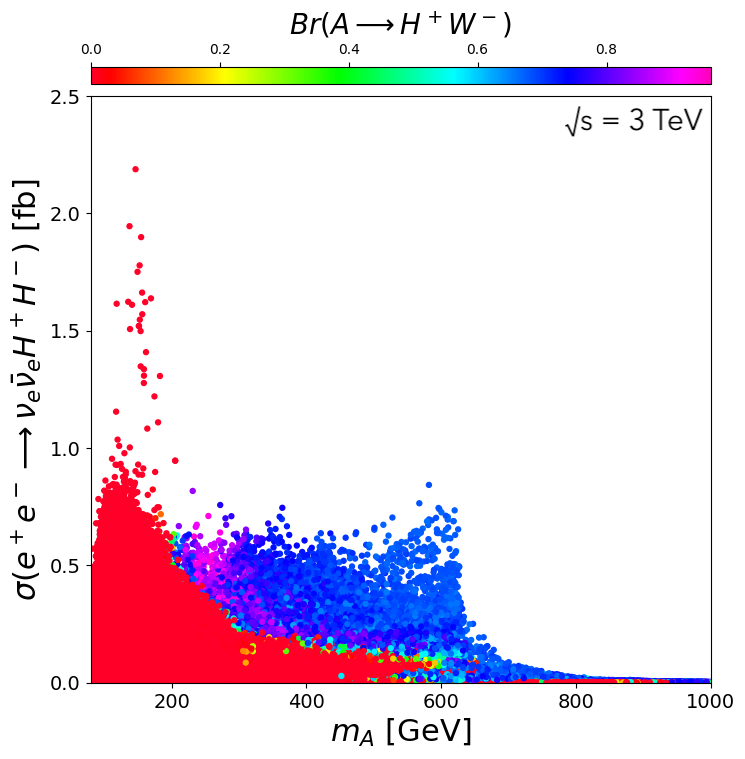}
 \includegraphics[scale=0.265]{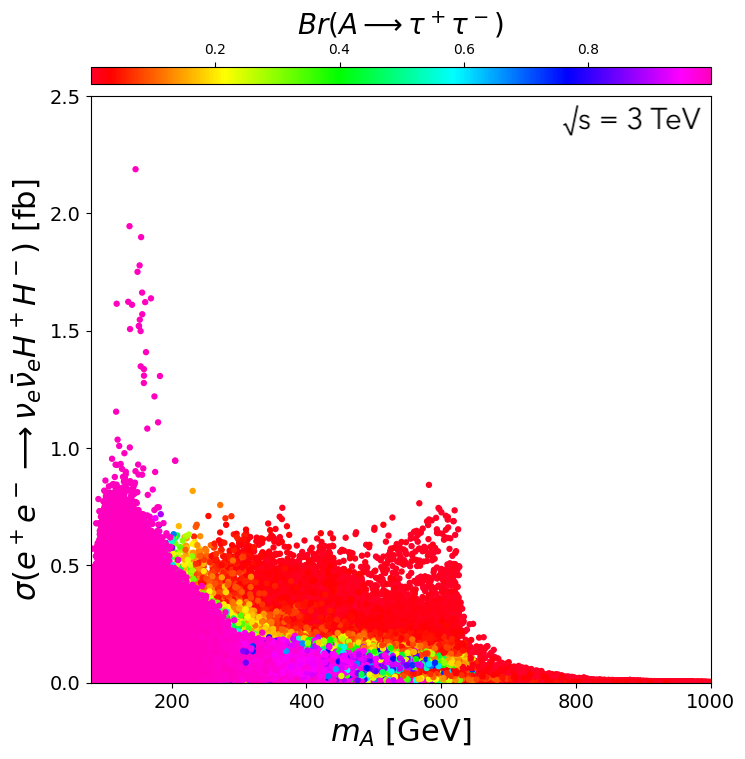}\\
 \includegraphics[scale=0.265]{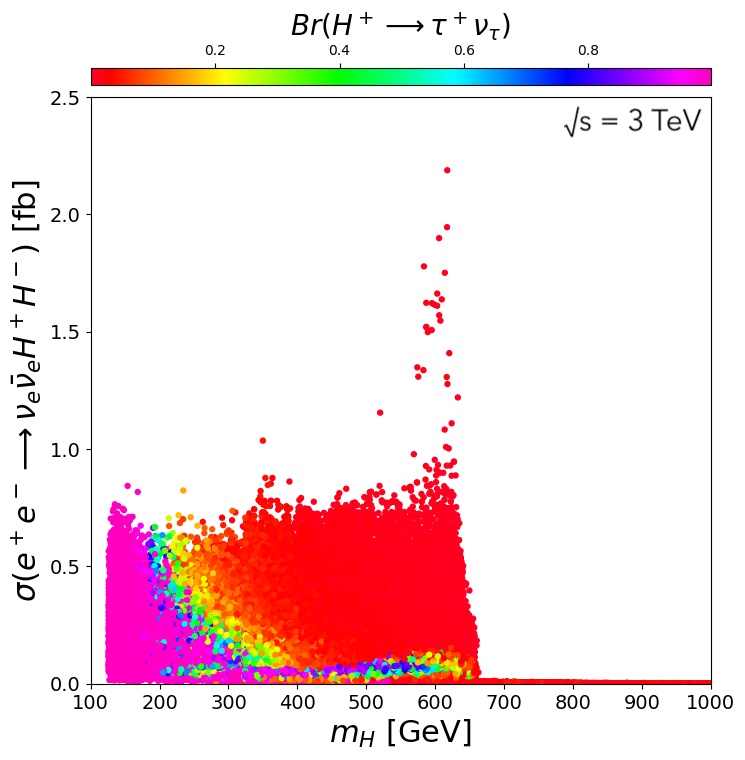}
 \includegraphics[scale=0.265]{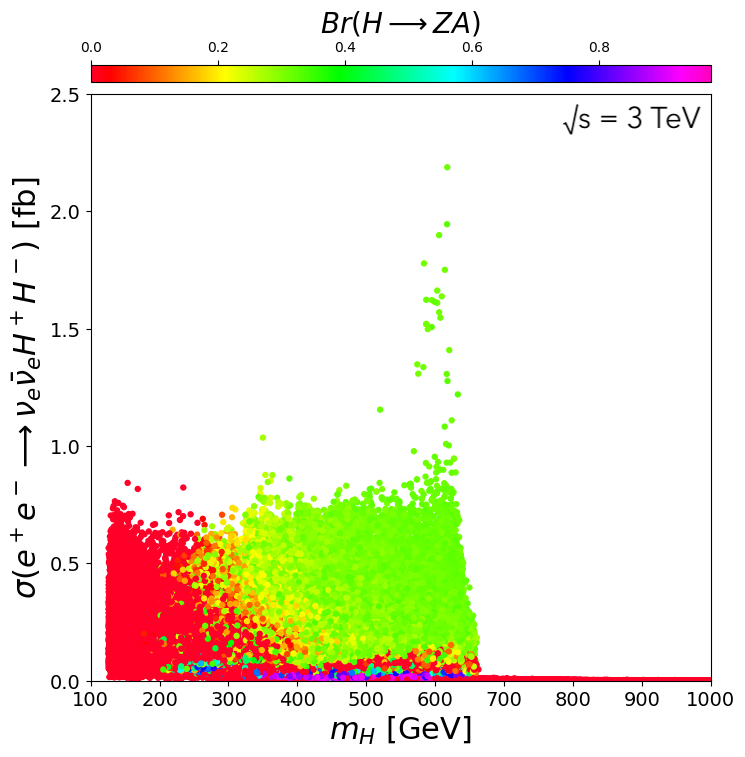}
 \caption{Production cross sections for $e^+e^- \to \nu_e \bar{\nu}_e H^+ H^-$ in Type X as a function of the model parameters, at the Compact Linear Collider (CLIC) with the center-of-mass energy of $\sqrt{s}=3$ TeV. The color legend presents one of the  following observables: $m_{H^\pm}$, $Br(H^+ \to \tau^+\bar{\nu}_\tau)$, $Br(H \to H^\pm W^\mp,\, Z A)$ and $Br(A \to ZH,\,H^\pm W^\mp,\,\tau^+\tau^-)$. All the regions are consistent with theoretical and experimental constraints.}
 		\label{fig1}
 	\end{center}
 \end{figure}
 
 The production cross sections for  3 TeV are depicted in Fig.\ref{cs_results_wh}.  
 The cross section for $\mu^+\mu^- \to W^\pm H^\mp$ is
 slightly lower than  $\mu^+\mu^- \to H^\pm H^\mp$.
 The difference could be attributed to the wider range of
 kinematics inherent in this process at a given $\sqrt{s}$. As expected the cross section  raise and is of
 the order of 0.63 fb for large value of $\tan\beta$. Also, one can see a large
 $Br(A \to H^{\pm}W^{\mp})$ effect from $\mu^+\mu^- \to A^*\to W^\pm H^\mp$,
 which could enhance the cross section while $Br(H\to W^\pm H^\mp)$ is
 rather small of the order $10^{-3}$. The competition between $H\to hh$, $H\to \tau^+ \tau^-$, and $H\to ZA$
 which makes $Br(H\to W^\pm H^\mp)$ rather suppressed.  One can see
 from the figures, the maximum cross section for $\mu^+\mu^- \to W^\pm
 H^\mp$ is could reach more that 0.63~fb.
 
\begin{itemize}
\item $e^+e^- \to \nu_e \bar{\nu}_e H^+ H^-$
\end{itemize}
Fig.\ref{fig1} illustrates our findings regarding the associated pairwise production of di-charged Higgs bosons with $\nu_e$, highlighting a strong dependence of the cross section on the model parameters. Hence, a significant variation in the production cross-section may occur, especially when the charged Higgs boson is relatively light and decays mainly to $\tau^{\pm}\bar{\nu}_\tau$. The CP-even $H$-exchange, as depicted in Fig.\ref{fig:diag-VBF-HpHm}-(b) or Fig.\ref{fig:diag-VBF-HpHm}-(d), contributes significantly for $m_H$ around 600 GeV. Conversely, the CP-odd Higgs boson, $A$, may both improve the cross section and give rise to $\tau^+\tau^-$ for $m_A \le 200$ GeV. However, for the muon collider context, while the new channels we highlighted in our study alter the landscape of potential production mechanisms, the impact of VBF processes warrants further investigation. Moving forward, we aim to extend our analysis to include a more comprehensive evaluation of VBF contributions in both $e^+e^-$ and muon collider setups, thereby enhancing our understanding of charged Higgs boson production in these environments.

%%=================== benchmarking h_SM
{\renewcommand{\arraystretch}{0.2} %donne la distance entre les lignes%
{\setlength{\tabcolsep}{0.1cm} %donne la distance entre les collones%
\begin{table}[hb]
\centering
\begin{tabular}{|c|c|c|c|c|c|c|c|c|c|}\hline
\\	
Parameters & $m_h$& $m_H$ & $m_A$ & $m_{H^\pm}$ & $\sin(\beta-\alpha)$ & $\tan\beta$&$m_{12}^2$ (GeV$^2$)&$\sigma^{3 TeV}_{Muon}$ (fb)&$\sigma^{14 TeV}_{LHC}$ (fb)\\  \\\hline
\\
\multicolumn{10}{|c|}{\small processes: charged Higgs pair production}\\ \\\hline
\\				
\text{BP1} & $125.09$ & $598.13$ & $148.36$ & $119.1$ & $0.9977$ & $30$ &$11873.03$&3.001&117.6 \\ \\ \hline
\\
\text{BP2} & $125.09$ & $509.68$ & $102.52$ & $95.11$ & $0.9994$ & $60$&$4328.42$&3.15&258.1\\ \\
\hline
\\
\multicolumn{10}{|c|}{\small processes: associate production }\\ \\\hline	
\\	
\text{BP1} & $125.09$ & $801.87$ & $973.69$ & $980.04$ & $0.9999$ & $12$&$53003.52$&0.048&5.619$\times 10^{-3}$\\ \\\hline
\\	
\text{BP2} & $125.09$ & $737.76$ & $904.88$ & $946.1$ & $0.9999$ & $30$&$18086.76$&0.304&1.062$\times 10^{-3}$\\ \\ \hline
\\
\text{BP3} & $125.09$ & $509.68$ & $102.52$ & $95.11$ & $0.9994$ & $60$&$4328.42$&1.1&12.37\\ \\\hline
\end{tabular}
\caption{The benchmark points (BPs) selected in Type X of the 2HDM
for differential cross sections.}
\label{Table:BPs}
\end{table}

 %===================
\begin{figure}[htb!]
\centering
\includegraphics[width=0.412\textwidth]{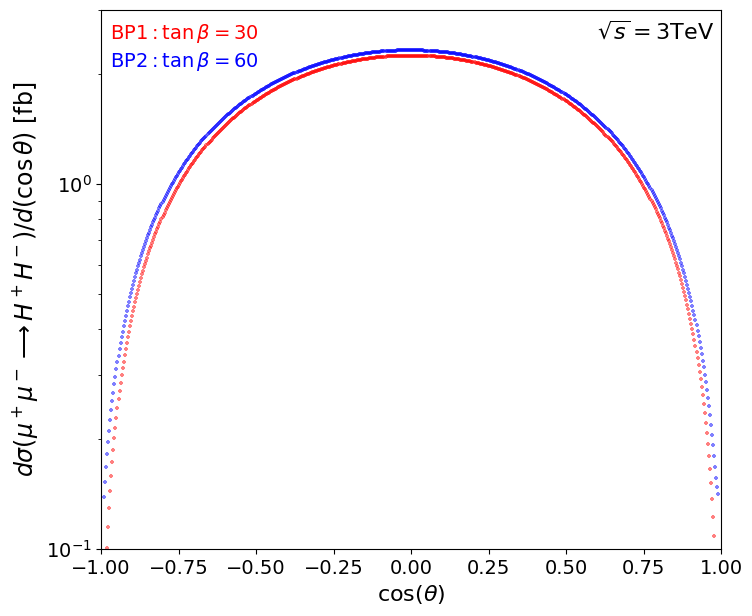}
\includegraphics[width=0.412\textwidth]{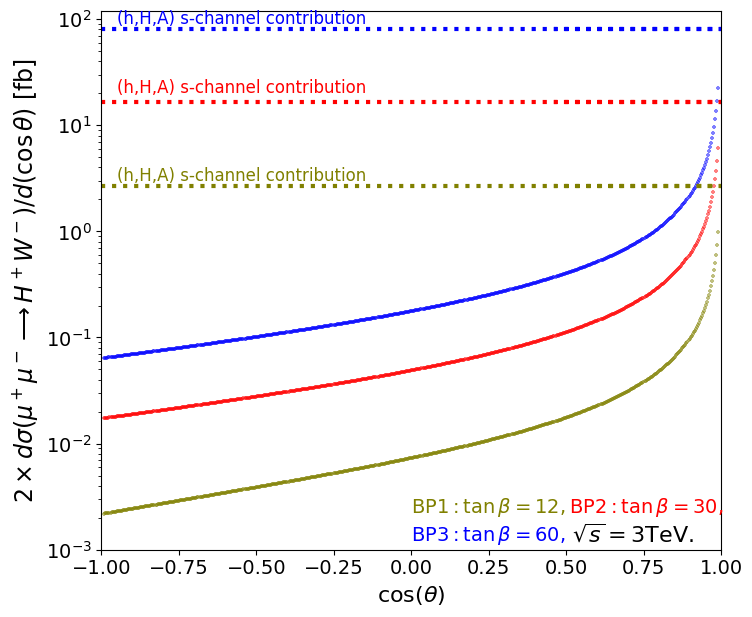}\\
\includegraphics[width=0.425\textwidth]{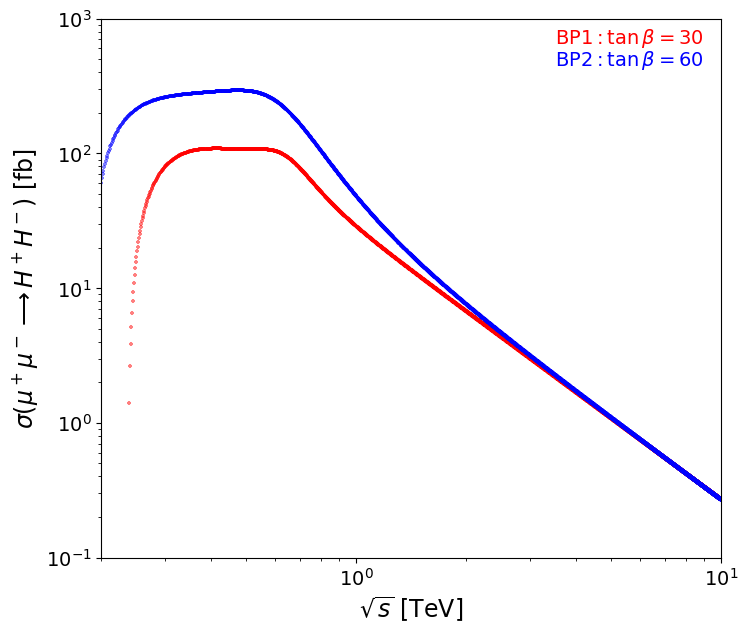}
\includegraphics[width=0.425\textwidth]{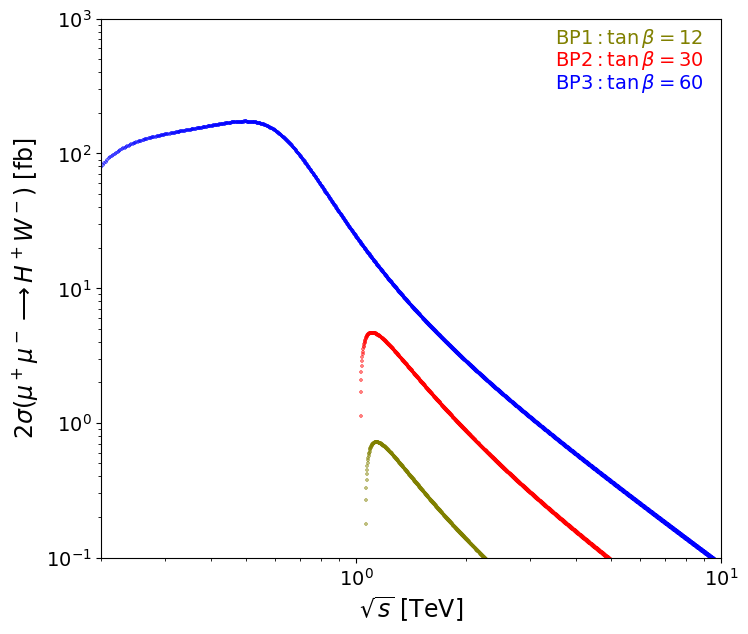}
\caption{Upper panel: angular distribution for $\mu^+\mu^- \to H^\pm H^\mp$ (left)
  and $\mu^+\mu^- \to H^\pm W^\mp$ (right) as a function of
  $\cos\theta$ at $\sqrt{s}=3$ TeV for the two benchmark points.
  Lower panel: total cross section for $\mu^+\mu^- \to H^\pm H^\mp$ (left)
  and $\mu^+\mu^- \to H^\pm W^\mp$ (right) as a function of $\sqrt{s}$ for
  the selected benchmark points. Note that $e^+ e^- \to H^\pm W^\mp$ is
  loop suppressed.}
\label{diff_cs_results}
\end{figure}

We illustrate the angular distribution $d\sigma/d\cos\theta$ for both
processes in Fig.\ref{diff_cs_results}(a) and (b) for
a selected benchmark points: $\tan\beta=12$, $\tan\beta=30$ and $\tan\beta=60$ 
at $\sqrt{s}=3$ TeV.
It is evident that for $H^+H^-$ pair production,
the distribution remains symmetric about $\cos\theta=0$. Such a
distribution depends solely on the exchanges of $\gamma$ and $Z$
in the $s$-channel. For the $s$-channel $h$ and $H$ contributions,
the angular distribution tends to be flat due to the absence of
scattering-angle dependence. Evidently, the angular distribution at
the muon collider is the same as for $e^+e^-\to H^+H^-$, except that
it is slightly shifted by the contributions of the scalar $h$ and $H$
exchanges,
which can nevertheless significantly improve the cross-section
production by several orders of magnitude for $\tan\beta=60$.

In Fig.\ref{diff_cs_results}(b), we display the differential
cross-section for $\mu^+\mu^- \to H^+ W^-$. This includes both the flat
differential distribution of the $s$-channel contributions $\mu^+ \mu^-
\to h^*, H^*, A^* \to W^\pm H^\mp$, as well as the total angular
distribution from both the $s$-channel and $t$-channel
contributions. It is clear that destructive interference occurs
between $s$- and $t$-channel contributions,
leading to suppressed total cross sections.
Notably, an enhancement near the forward direction
($\cos\theta\approx1$) can be seen due to the $t$-channel singularity.
 
Figures~\ref{diff_cs_results}(c) and (d) present the total cross sections
for both processes at $e^+e^-$ and $\mu^+\mu^-$ colliders
as a function of $\sqrt{s}$ for various values of
$\tan\beta$. For both processes, we select benchmark points to
highlight the resonance effect from the $s$-channel $H$ contribution.
In the case of $\sigma(\mu^+\mu^-\to H^+ H^-)$, such resonance effects
is amplified with large $\tan\beta$. At higher energy $\sqrt{s}>2$ TeV, the
cross section for $\sigma(\mu^+\mu^-\to H^+ H^-)$ exhibits no
$\tan\beta$ dependence and is fully dominated by the $\gamma$ and $Z$
exchanges. For comparison we also show the angular distribution
for $e^+e^-\to H^+ H^-$ at $\sqrt{s}=1$ TeV and for charged Higgs
mass as specified for $\text{BP}_{1,2}$.
Similarly, in the case of $\sigma(\mu^+\mu^-\to H^\pm W^\mp)$,
one can see the resonance effect from the $H$ exchange. Due
to the destructive interference between the $s$-channel and $t$-channel
contributions, one can see a quick drop of the cross section as a
function of $\sqrt{s}$.
 
Implicitly, the $\sigma(\mu^+\mu^-\to H^\pm W^\mp)$ depends both
on the energy of the muon collider as well as $\tan\beta$, unlike the
$H^+ H^-$ production mode where the total cross section drastically
decreases for high $\sqrt{s}$ values and becomes insensitive to
$\tan\beta$. Furthermore, contrary to $\sigma(\mu^+\mu^-\to H^+ H^-)$,
the fact that the amplitude of $\mu^+\mu^-\to H^\pm W^\mp$ proportional
to $\tan\beta$ results in a clear shift between $\text{BP}_{1}$ for
$\tan\beta=30$ and $\text{BP}_{2}$ for $\tan\beta=60$ as can be seen
from Fig.\ref{diff_cs_results}(d). Note that $e^+ e^- \to H^\pm W^\mp$ is loop suppressed and thus 
not shown here.
  
 At the end of this section, we stress that all the results displayed above are for 2HDM type X.  
 Since all of the neutral and charged Higgs couplings to the muons involved are the same in type II and type X, 
 we would get the same numerical result for the same set of parameters for 2HDM type II. The same statement is true for 2HDM types I and Y.
The parameter space for the 2HDM type II is constrained in two ways. 
The first comes from B-Physic observables, as noted by \cite{Haller:2018nnx,HFLAV:2016hnz}. 
The most demanding requirement comes from $\bar{B}\to X_s \gamma$, 
which demands that the charged Higgs boson to be heavier than 680 GeV \cite{Hermann:2012fc,Misiak:2017bgg,Misiak:2020vlo}.
The second restriction is caused by LHC data which requires $\tan\beta \leq 12$. As a result, the production of charged Higgs pairs is only possible for large center of mass energies $\sqrt{s}\geq 1360$ GeV, and in such circumstance, the s-channel neutral Higgs contribution is largely suppressed.  In this scenario, the cross section would resemble $\sigma (e^+ e^-\to H^+ H^-)$ given  in Eq.(\ref{eq:tot_cs_ee_hphm}). In the case of  $\mu^+\mu^- \to  W^\pm H^\mp$ in the 2HDM type X,  for charged Higgs heavier than 680 GeV and $\tan\beta <12$, at $\sqrt{s}=1$ TeV the maximum cross section one can get is about
a few fb. This cross section is even reduced for $\sqrt{s}\geq 1$ TeV.\\
However, the resulting cross section for $\mu^+\mu^- \to H^+H^-$ in the 2HDM type I and Y does not yield any meaningful results. This outcome results from the fact that the couplings $\xi_{H}^\ell=s_{\alpha}/s_{\beta} \propto 1/\tan\beta$ is suppressed 
for large $\tan\beta$ and the coupling $\xi_h^{\ell}$ reduces to unity. Therefore, the primary component 
of the cross section arises from the $\mu^+\mu^- \to \gamma^* , Z^*  \to H^+H^-$ given  in Eq.(\ref{eq:tot_cs_ee_hphm}). 
The same thing can be said for $\sigma(\mu^+\mu^- \to H^\pm W^\mp)$ cross section which is proportional
to $1/\tan^2\beta$ at the large $\tan\beta$ limit and is therefore small.

 \section{Signal vs. Background Analysis}
\label{section5}
\subsection{Monte Carlo Toolchain}
\label{section3subsec4}

This study for future muon collider operating at $\sqrt{s}$= 3 TeV
focuses on the  final state, which consists of a pair of
tau-leptons plus missing energy through the decay mode $\mu^{+}
\mu^{-} \rightarrow H^{+} H^{-} \rightarrow \tau_h^+ \nu \tau_h^-
\nu$,  VBF channel and  $\mu^{+} \mu^{-} \rightarrow H^{+} W^{-}
\rightarrow W^{+} W^{-}h \rightarrow \tau_h^+ \nu \tau_h^-
\nu$.  For each
channel, we adopt the benchmarks set in Table.\ref{Bp1}.

\begin{table}[h]
	\setlength{\tabcolsep}{7pt}
	\renewcommand{\arraystretch}{1.2}
	\centering
	\begin{tabular}{|c|c|c|c|c|c|c|c|c|}       
		\hline \hline 
		&signal& $m_h$  & $m_H$ & $m_A$ & $m_{H^{\pm}}$  &  $\tan \beta$  &  $\sin (\beta -\alpha)$ & $m_{12}^2$ \\\hline\hline

		\text{BP1}&$ \tau_h^+ \nu \tau_h^- \nu$ & 125.09 & 590.5& 234.04 & 207.28 & 44.63 & 0.9990 & 7808.71 \\ \hline
		\text{BP2}&VBF & 125.09 & 622.39 & 267.96 & 248.14 & 32.18 & 0.9981 & 12025.25 \\ \hline
		\text{BP3}&$ \tau_h^+ \nu \tau_h^- \nu$ & 125.09 & 608.51 & 166.7 & 162.7 & 28.57 & 0.9977 & 12942.8 \\ \hline
		\hline \hline
	\end{tabular}
	\caption{The description of our BPs.}\label{Bp1}
\end{table} 

To simulate the signal events, we generate the parton-level processes
using \texttt{MadGraph5\_aMC\_v3.4.1} \cite{Alwall:2014hca}. 
Event samples are then interfaced with \texttt{Pythia-8.20}
\cite{Sjostrand:2007gs} for fragmentation and showering, and
subsequently processed through the \texttt{Delphes-3.4.5}
\cite{deFavereau:2013fsa} for detector simulation, where we use the muon
collider Detector TARGET model. Jets are clustered using the
anti-$k_t$ \cite{Cacciari:2008gp} algorithm through \texttt{Delphes} with
a jet radius $R = 0.5$. At the Delphes level, we first require that the
candidate for a b-jet should pass the minimal acceptance 
of $p_T >$ 20 GeV. Then we apply the b-tagging efficiency about
70$\%$ and the mistag rates of the charm or light quark jet as a
b-jet as a function of pseudorapidity and energy.
To assess the observability, we evaluate the statistical
significance (S) using the
formula:
\begin{eqnarray}
	S=\sqrt{ \mathcal{L}} \frac{\sigma _s}{\sqrt{\sigma _s+\sigma _b}} \;,
\end{eqnarray}
where $\sigma _s$ and $\sigma _b$ are the signal and background cross sections after all the cuts
and $\mathcal{L}$ is the integrated luminosity.

\subsection{$\mu^+\mu^- \to H^+H^-$ and VBF production,}
Production of a pair of charged-Higgs bosons is considered to
be one of the most challenging processes at the LHC. In
this subsection, we focus on pair production of charged Higgs at
future muon collider and VBF production with the final state $\tau^{+} \nu \tau^{-} \nu $ :
\[
\mu^{+} \mu^{-} \rightarrow H^{+} H^{-} \rightarrow  \tau_h^+ \nu \tau_h^- \nu
\]

To search the signals against the SM background, we present
the main SM backgrounds, which include top-pair production
$t\bar t$, diboson production $VV$=$WW$=$ZZ$, $Zjj$, and $Wjj $ listed
below.
\begin{itemize}
	\item $\mu^{+} \mu^{-} \rightarrow t\bar{t}$
	\item $\mu^{+} \mu^{-} \rightarrow VV$ 
	\item $\mu^{+} \mu^{-} \rightarrow Z/\gamma \ jj$ with $Z \rightarrow \tau \tau $ and $Z \rightarrow \nu \nu$
	\item $\mu^{+} \mu^{-} \rightarrow Wjj$ where one $\tau_h$ comes from W decay and the other $\tau_h$ from a jet misidentified as $\tau_h$.
\end{itemize}
To ensure that the parton-level events meet the required criteria,
we impose $p_T^{j} > 25$ and $|\eta_{j}| < 2.5$ on jets. However, at
Delphes level we employ the $\tau$-tagging efficiencies and the mistag
rates of a light jet as $\tau$ $P_{\tau \rightarrow \tau}$=0.85 and
$P_{j \rightarrow \tau }$=0.02, respectively.
Furthermore, we apply the charged lepton identification and
typical photon isolation criteria, where we
require
\begin{center}
	$I(P)= \frac{1}{p_{T}^{P}} \Sigma p_{T_{i}} < 0.01$   :
\end{center}

\begin{figure}[h]
\centering	
\includegraphics[width=0.43\textwidth]{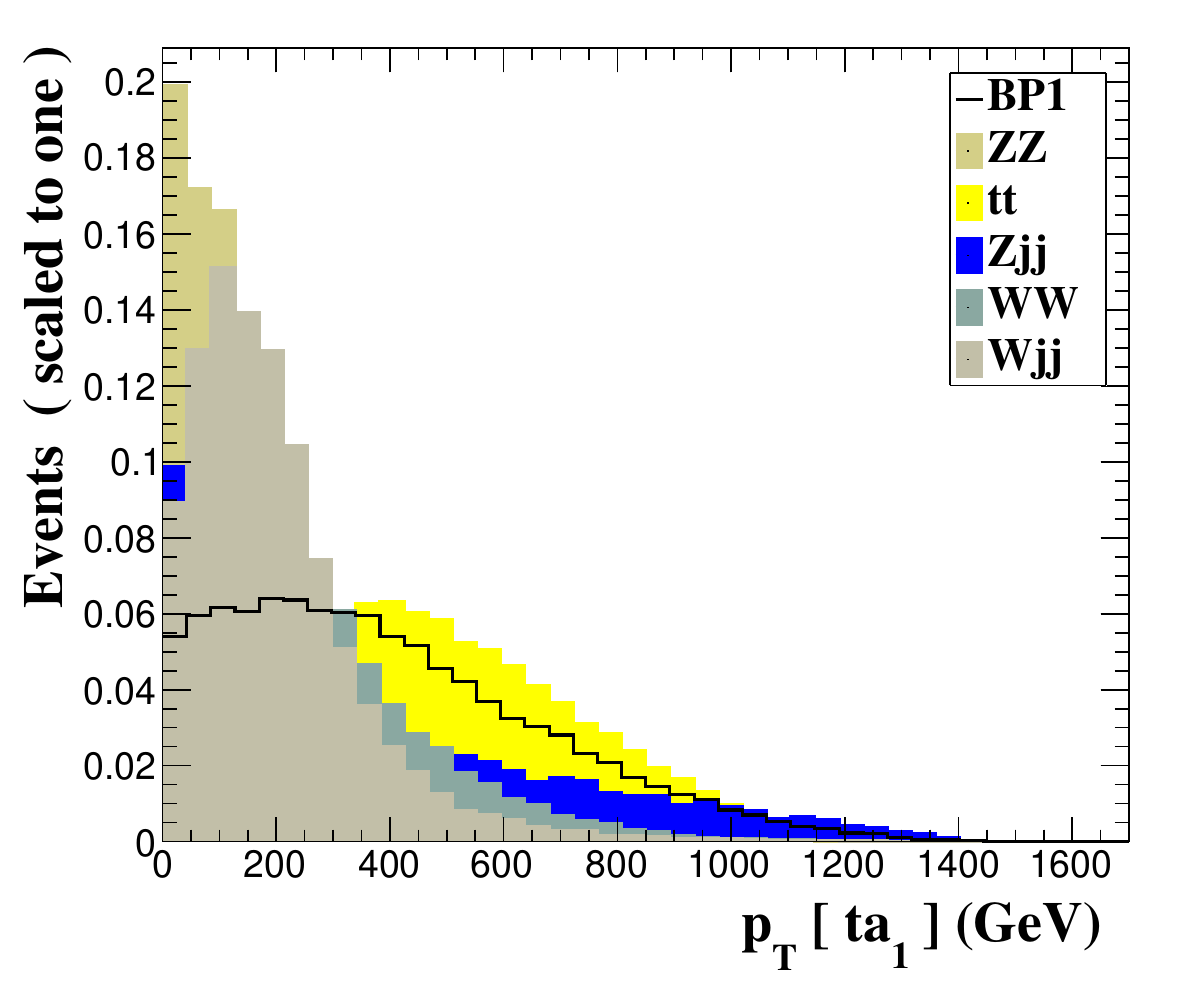}
\includegraphics[width=0.43\textwidth]{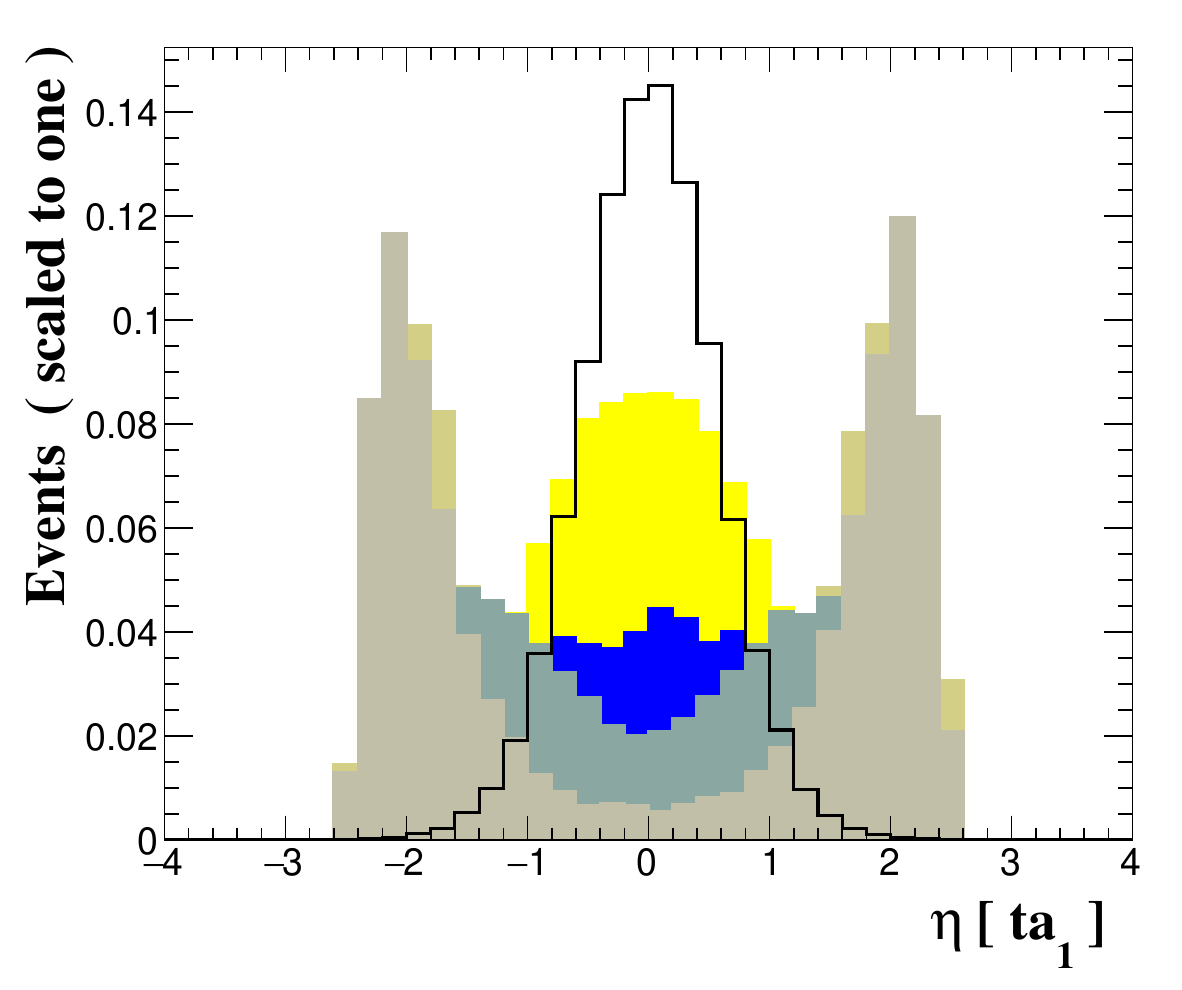}\\
\includegraphics[width=0.43\textwidth]{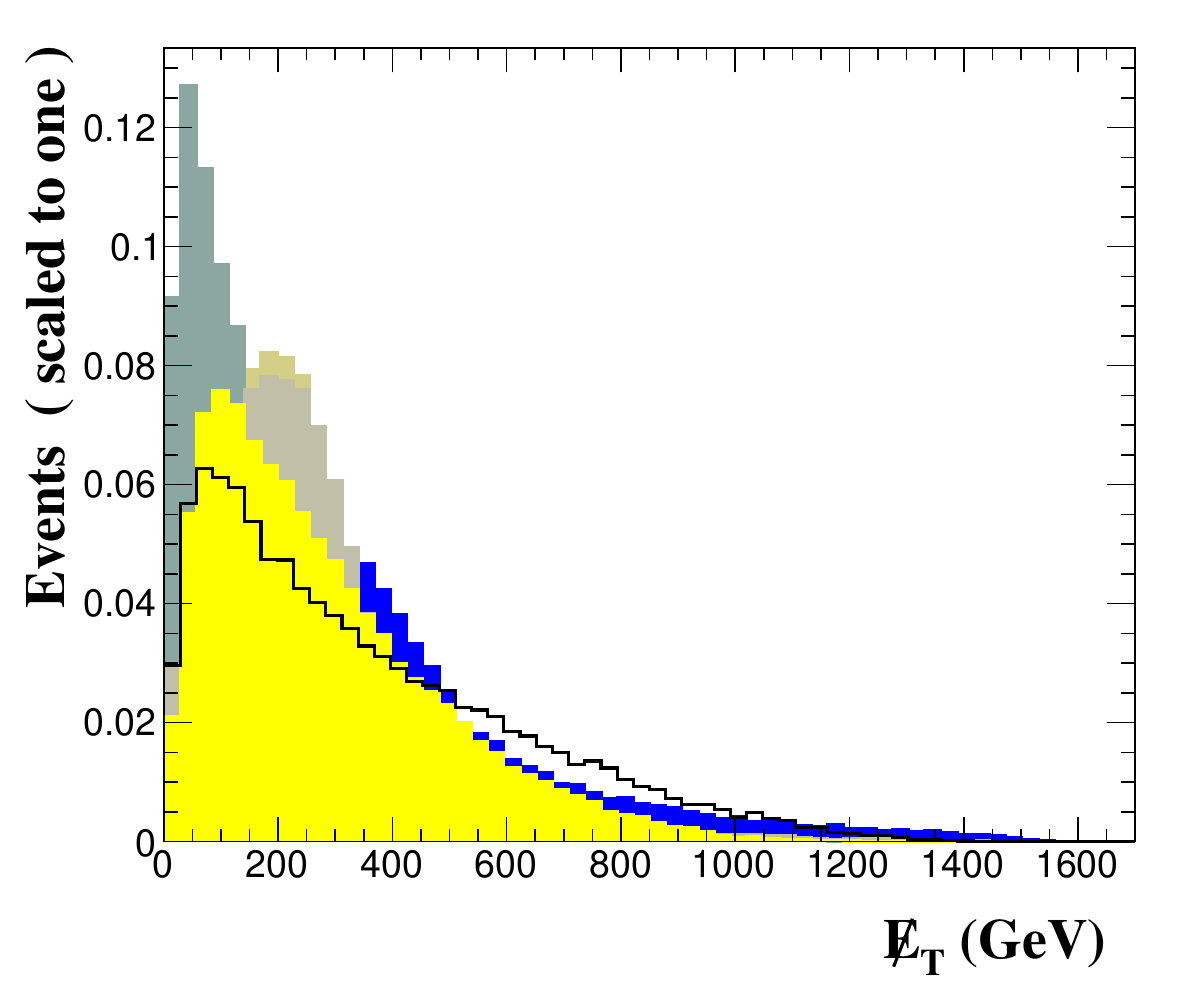}
\includegraphics[width=0.43\textwidth]{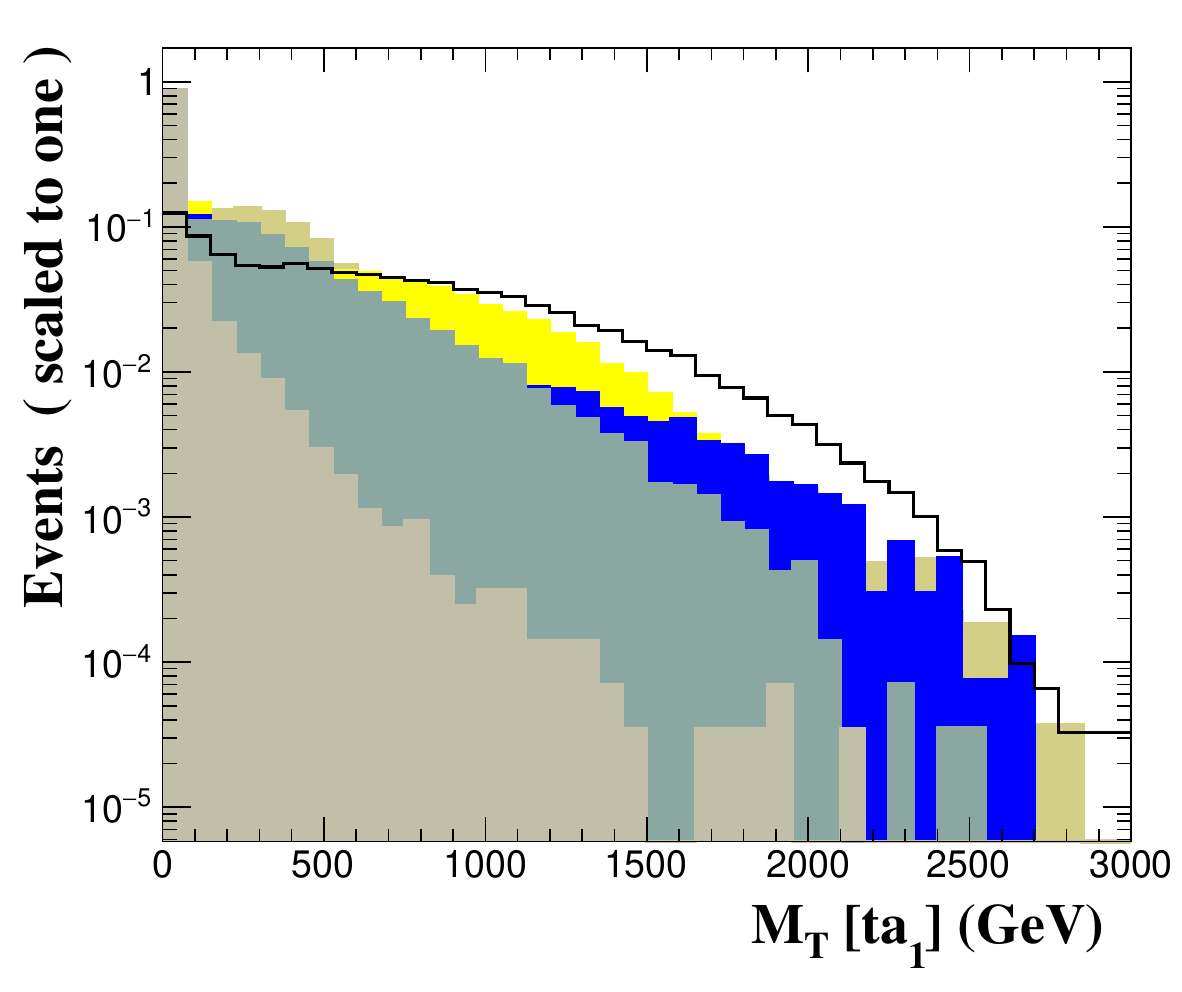}
\caption{Normalized kinematic distributions for the final state $\tau^+ \nu \tau^-  \bar{\nu}$: transverse momentum of the leading tau lepton (ordered in $p_T$) $p_T$[$\tau_1 $] (top-left panel), the pseudorapidity of the leading tau lepton $\eta$ [$\tau_1 $ ] (top-right panel), the transverse energy $E_T$ (lower left panel), and  the transverse mass $M_T$ [$\tau_1 $] (lower-right panel) at $\sqrt{s}=3$ TeV muon collider.}
\label{t-001}
\end{figure}

Fig.\ref{t-001} presents the distributions for the transverse momentum
of the leading tau lepton $p_{T}\ [\tau_{1}]$ (ordered by $p_T$),
the pseudorapidity of the leading tau $\eta\ [\tau_1]$, 
transverse mass $M_{T}\ [\tau_{1}]$, and the transverse energy
for the signal benchmark point $BP1$ and various
SM backgrounds at the 3 GeV muon collider.
In order to increase the significance of the signal, we have
established a cut-flow based on the behavior of the kinematic
distributions, as presented in Table.\ref{cutft}.
We start by limiting the number of b-quark N(b) $\leq 1$,
which is vital in discriminating the signal from the background.
Under this cut the $t\bar t$ background survived only 47$\%$ of
events without putting any impact on the signal events.
The first selection cut imposed is $P_{T}[\tau_{1}] >60 $ GeV  and $-0.5< \eta [\tau_{1}]<0.5$  going together for the detection of the tau leptons, under this cut significantly reduces a substantial portion of the background, particularly eliminating the majority of $Wjj$, $ZZ$, $WW$ and $Zjj$ events .
The next cut is the missing energy requirement $\slashed{E}_{T}> 500 $ GeV,
which removes about $53\%$ of the $tt$, $72\%$ of the $Zjj$,
64\% of $WW$  events,
while the survival rate for the signal is more than 43$\%$.
The lower-right panel (last selection cut) in Fig.\ref{t-001} indicates a 
transverse momentum $M_{T}[\tau_{1}] >600 $GeV cut would be useful, which
constitutes the ``Cut-3''. 
It removes about 94\% of $Wjj$ 
while 97$\%$ of the signal survives. For the VBF production, the cuts are totally different, as shown in Fig.\ref{fig77}. The achieved signal and background separation, as presented in Tab.\ref{cut4} for the signal and background, we impose number of b-quark N(b) $\leq 1$. The First selection cut applied is   $\slashed{E}_{T} < 180 $ GeV. Following, by  imposing  $E_{T} < 300 $ GeV, under this cut the  $t\bar{t}$ and $Wjj$ background events  are totally killed.

\begin{figure}[h]
\centering	
\includegraphics[width=0.43\textwidth]{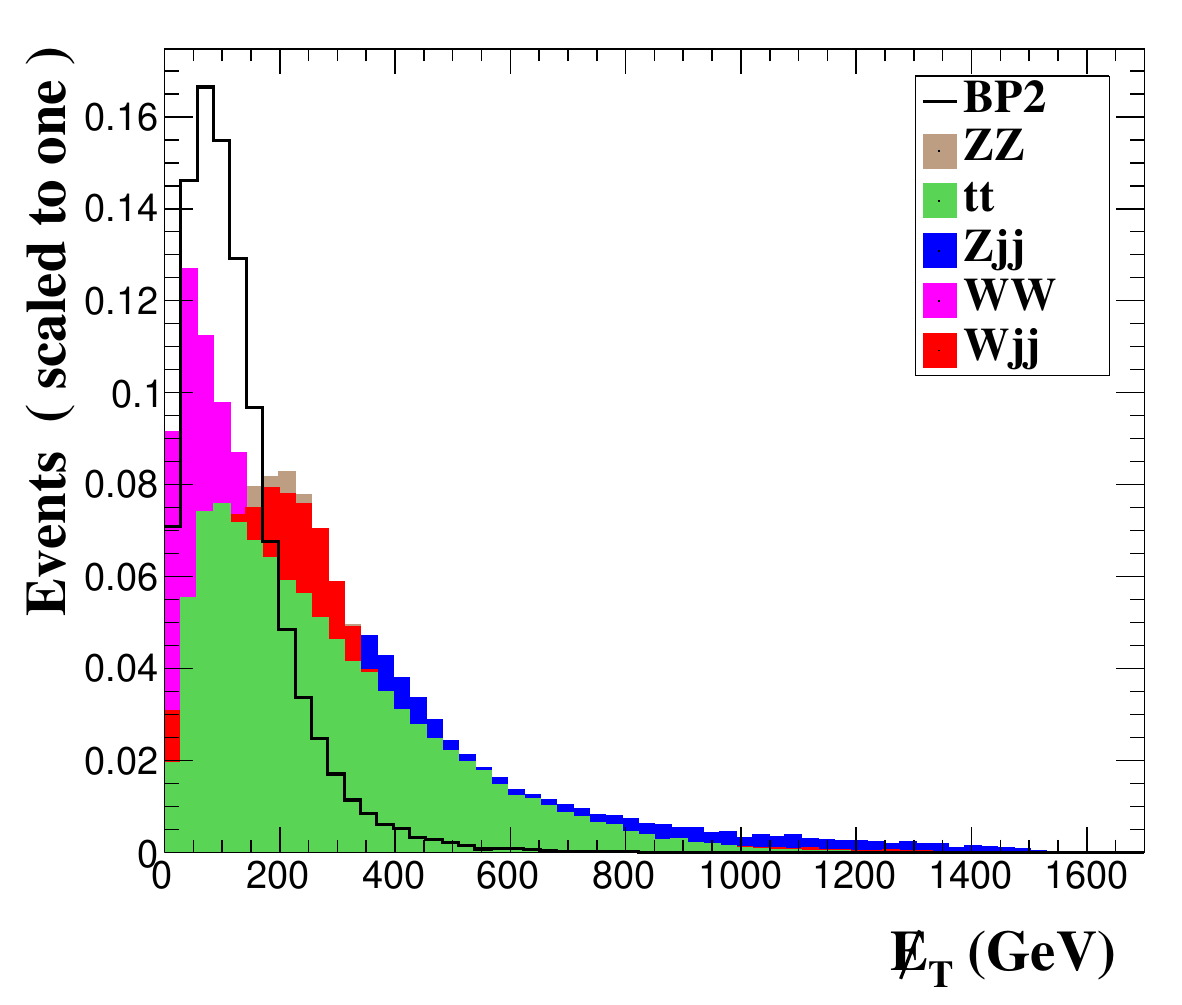}
\includegraphics[width=0.43\textwidth]{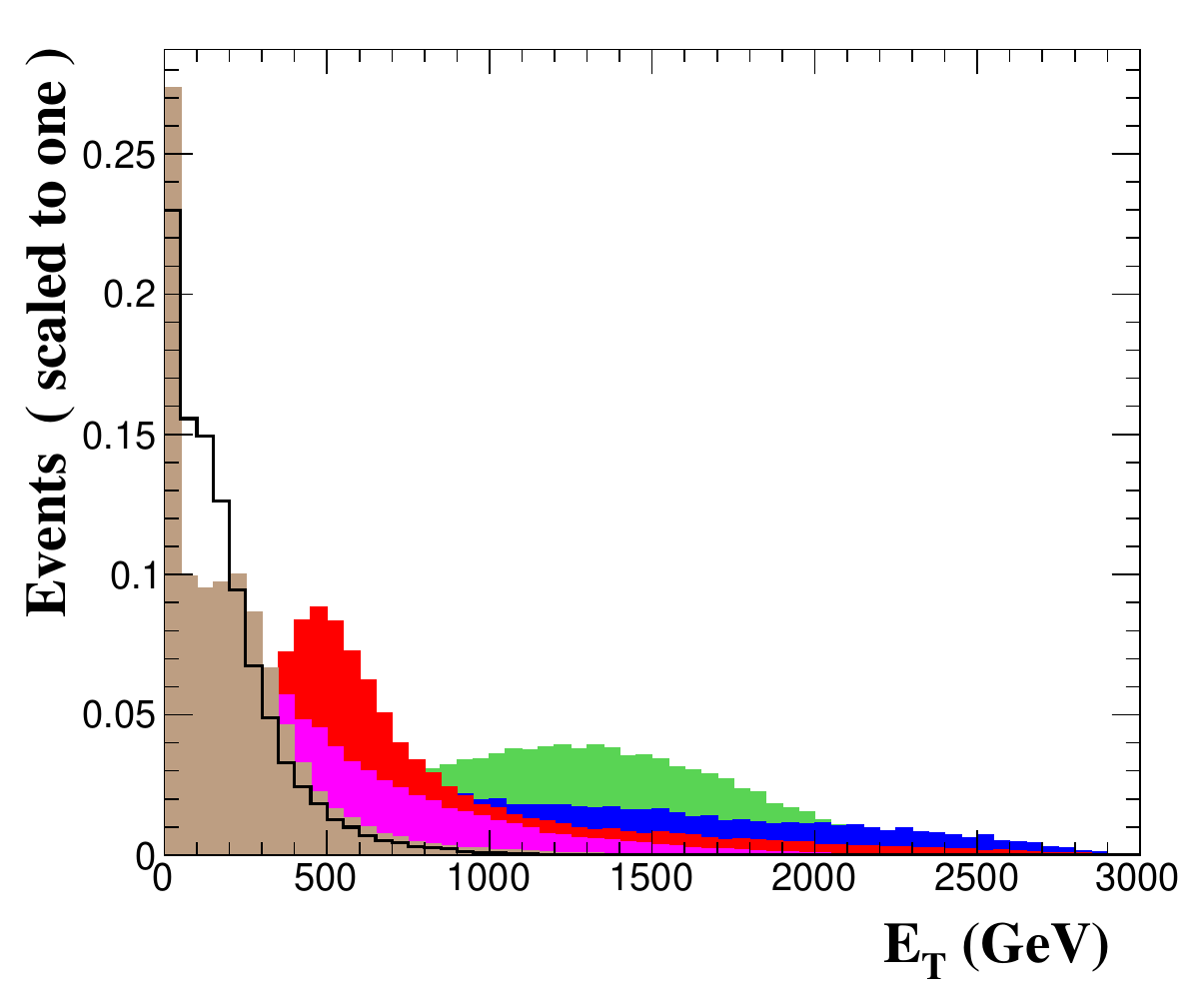}\\
\caption{Normalized kinematic distributions of the signal and backgrounds for the VBF production: the missing transverse energy $\slashed{E}_{T}$ (left panel) and the transverse energy $E_T$ (right panel) at $\sqrt{s}=$3 TeV muon collider.} \label{fig77}
\end{figure}

\begin{table}[h]
	\centering
	\setlength{\tabcolsep}{47pt}
	\renewcommand{\arraystretch}{1.2}
	\begin{adjustbox}{max width=\textwidth}
		\begin{tabular}{cc} 
			\hline  \hline
			Cuts & Definition \\  
			\hline \hline	
			trigger & $N(b)\leq 1$\\  	
			\hline
			Cut-1 & $P_{T}[\tau_{1}] >60 $ GeV  and $-0.5< \eta [\tau_{1}]<0.5$  \\  
			\hline
			Cut-2  &  $ \slashed{E_{T}} > 600$ GeV    \\  
			\hline
			Cut-3 &     $  M_T[\tau_{1}] > 600 $ GeV \\
			\hline \hline
		\end{tabular}
	\end{adjustbox}		
	\caption{A set of cuts used in the signal-background analysis of
		$\mu^+ \mu^- \to H^+ H^- \to \tau^+_h \nu \tau^-_h \nu$ 
		at $\sqrt{s}=3$ TeV.}
	\label{oo}
\end{table}

\begin{table}[h]
	\centering
	\setlength{\tabcolsep}{67pt}
	\renewcommand{\arraystretch}{1.2}
	\begin{adjustbox}{max width=\textwidth}
		\begin{tabular}{cc} 
			\hline  \hline
			Cuts & Definition \\  
			\hline \hline	
			trigger & $N(b)\leq 1$\\  	
			\hline
			Cut-1 & $ \slashed{E_{T}} < 180$ GeV \\  
			\hline
			Cut-2  &  $ E_{T} < 300$ GeV    \\  
			\hline \hline
		\end{tabular}
	\end{adjustbox}		
	\caption{A set of cuts used in the signal-background analysis of
		Vector Boson Fusion (VBF) 
		at $\sqrt{s}=3$ TeV.}
	\label{oo}
\end{table}

\begin{table}
	\centering
	\setlength{\tabcolsep}{5.pt}
	\renewcommand{\arraystretch}{1}
	\begin{tabular}{p{3cm}<{\centering}  p{1.2cm}<{\centering} p{1.4cm}<{\centering}p{1.4cm}<{\centering}  p{1.6cm}<{\centering} p{1.6cm}<{\centering} p{1.4cm}<{\centering} p{1.4cm}<{\centering} p{2cm}<{\centering} p{2cm}<{\centering}  p{2cm}<{\centering}p{0cm}<{\centering}}
		\hline\hline
		\multirow{2}{*}{Cuts$\ \ \ \ \ \ \ \ \ \ \ \ \ \ \ $}& \multicolumn{1}{c}{Signal }& \multicolumn{3}{c}{~~Backgrounds}&  \\ \cline{2-2}  \cline{4-8}
		&  $\text{BP1}$ &&$t\bar t $& $WW$  & $ZZ$ & $Wjj$&$Z/\gamma \ jj$\\
		\hline\hline
		Basic cut $\ \ \ \ \ \ \ \ $ &2.98  &&0.2 &1.54&0.014& 7.04& 0.74 \\
		Tagger $\ \ \ \ \ \ \ \ \ \ \ $& 2.97 &&0.09 &1.53 &0.014 &6.39 & 0.68\\
		Cut-1$\ \ \ \ \ \ \ \ \ \ \ \ \ \ $&0.72   && 0.04 &0.052 &0.0001&0.048 &0.02\\
		Cut-2$\ \ \ \ \ \ \ \ \ \ \ \ \ \ $&0.31 && 0.009&0.018 &9.$10^{-5}$ &0.021 &0.006\\
		Cut-3$\ \ \ \ \ \ \ \ \ \ \ \ \ \ $& 0.30  && 0.008 &0.017 &9.$10^{-5}$  & 0.0014 &0.005\\
		Total efficiencies & $10.4\%$  && 4$\%$ & 1.1$\%$ & 0.6$\%$ & $2. 10^{-2}
		\%$ & $0.6\%$ \\
		\hline\hline
	\end{tabular}
	\caption{The cut flow of the cross sections (in fb) for the
		signal and SM backgrounds at $\sqrt{s}=3$ TeV muon collider with  our typical $\text{BP1}$ \label{cutft}.}
\end{table}

\begin{table}
	\centering
	\setlength{\tabcolsep}{5.pt}
	\renewcommand{\arraystretch}{1}
	\begin{tabular}{p{3cm}<{\centering}  p{1.2cm}<{\centering} p{1.4cm}<{\centering}p{1.4cm}<{\centering}  p{1.6cm}<{\centering} p{1.6cm}<{\centering} p{1.4cm}<{\centering} p{1.4cm}<{\centering} p{2cm}<{\centering} p{2cm}<{\centering}  p{2cm}<{\centering}p{0cm}<{\centering}}
		\hline\hline
		\multirow{2}{*}{Cuts$\ \ \ \ \ \ \ \ \ \ \ \ \ \ \ $}& \multicolumn{1}{c}{Signal }& \multicolumn{3}{c}{~~Backgrounds}&  \\ \cline{2-2}  \cline{4-8}
		&  $\text{BP2}$ &&$t\bar t $& $WW$  & $ZZ$ & $Wjj$&$Z/\gamma \ jj$\\
		\hline\hline
		Basic cut $\ \ \ \ \ \ \ \ $ &2.35 &&0.2 &1.54&0.014& 7.01& 0.743 \\
		Tagger $\ \ \ \ \ \ \ \ \ \ \ $& 2.34 &&0.09 &1.53 &0.014 &6.38 & 0.68\\
		Cut-1$\ \ \ \ \ \ \ \ \ \ \ \ \ \ $&1.85   && 0.03 &0.94 &0.005&2.44&0.2\\
		Cut-2$\ \ \ \ \ \ \ \ \ \ \ \ \ \ $&1.66 && 0.0009&0.52 &0.005 &0.23 &0.024\\
		Total efficiencies & $70\%$  && 0.45$\%$ &  33$\%$ & $35\%$ & 3.2$\%$ &$3.2\%$ \\
		\hline\hline
	\end{tabular}
	\caption{The cut flow of the cross sections (in fb) for the
		VBF signal and SM backgrounds at $\sqrt{s}=3$ TeV muon collider with  our typical $\text{BP2}$\label{cut4}.}
\end{table}

\begin{figure}[h]
\centering
\includegraphics[width=0.4\textwidth]{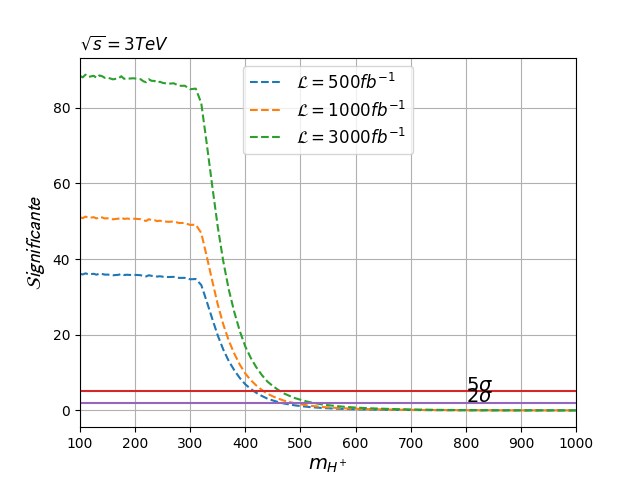}
\includegraphics[width=0.4\textwidth]{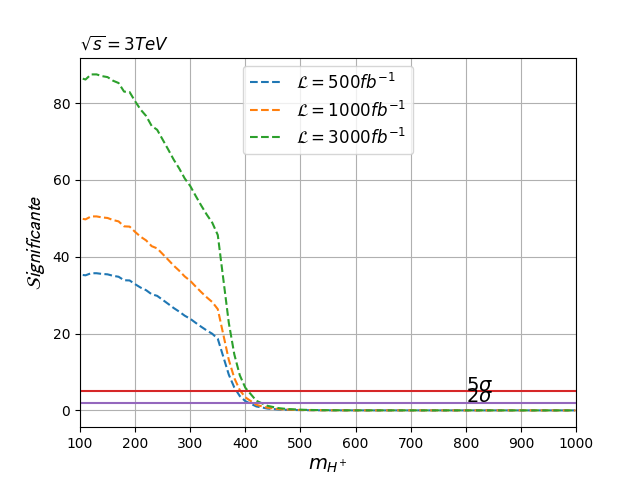}
\caption{The significance obtained for the process $\mu^+ \mu^- \to H^+ H^- \to \tau^+ \nu \tau^- \nu$ (left panel) and VBF production (right panel) versus the charged-Higgs mass at 3 TeV muon collider with integrated luminosities of 500 fb$^{-1}$ , 1000 fb$^{-1}$ and 3000 fb$^{-1}$. The 2$\sigma$ and 5$\sigma$ significance levels are also indicated.}
\label{fig88}
\end{figure}  

From Fig.\ref{fig88} (left panel), which presents the significance ($S$) as a function
of charged-Higgs boson mass $m_{H^{+}}$ of the process $\mu^+ \mu^- \to H^+ H^-$ at $\sqrt{s}=3$ TeV for
various integrated luminosities $\mathcal{L}= 500 fb^{-1}$ , 1000 fb$^{-1}$ and 3000 fb$^{-1}$, where the red solid line
denotes the 5$\sigma$ discovery and the purple solid line denotes
the 2$\sigma$ exclusion.  With the mere integrated luminosities of 500 fb$^{-1}$ , 1000 fb$^{-1}$ and 3000 fb$^{-1}$, 
the outlook for significance seems highly 
promising. The prospects for significance appear highly promising. the region $m_{H^{+}}\in$ [100GeV, 450GeV] can be discovered. In the right panel, we show the significance for the VBF channels at $e^+e^- $ machines. Furthermore, the discovery 5$\sigma$ level achieved at  various integrated luminosities $\mathcal{L}=500 fb^{-1}$ , 1000 fb$^{-1}$ and 3000 fb$^{-1}$ in the range masses  $m_{H^{+}}\in$ [100GeV, 405GeV]. In summary, we have demonstrated the feasibility of achieving highly favorable signal identification over the standard model backgrounds in the context of charged Higgs production, considering both direct $\mu^{+}\mu^{-}$ and the vector boson fusion (VBF) channels.
	
\subsection{$\mu^+\mu^- \to H^{\pm} W^{\mp}$}
The most important production channel for the charged Higgs is the associated production with a $W$ boson, which merits particular attention. The Feynman diagrams for the process are modeled in four separate subprocesses, as illustrated in Fig.\ref{diag:2mu2HpWm}, each of which contributes to the $H^{\pm}W^{\pm}$ production cross section. We explore the following final state
\[ \mu^{+} \mu^{-} \rightarrow H^{\pm} W^{\mp} \rightarrow  \tau^+ \nu \tau^- \nu \]
	
The corresponding SM backgrounds that can mimic our final state are listed above.
		
\begin{figure}[h]
\centering	
\includegraphics[width=0.43\textwidth]{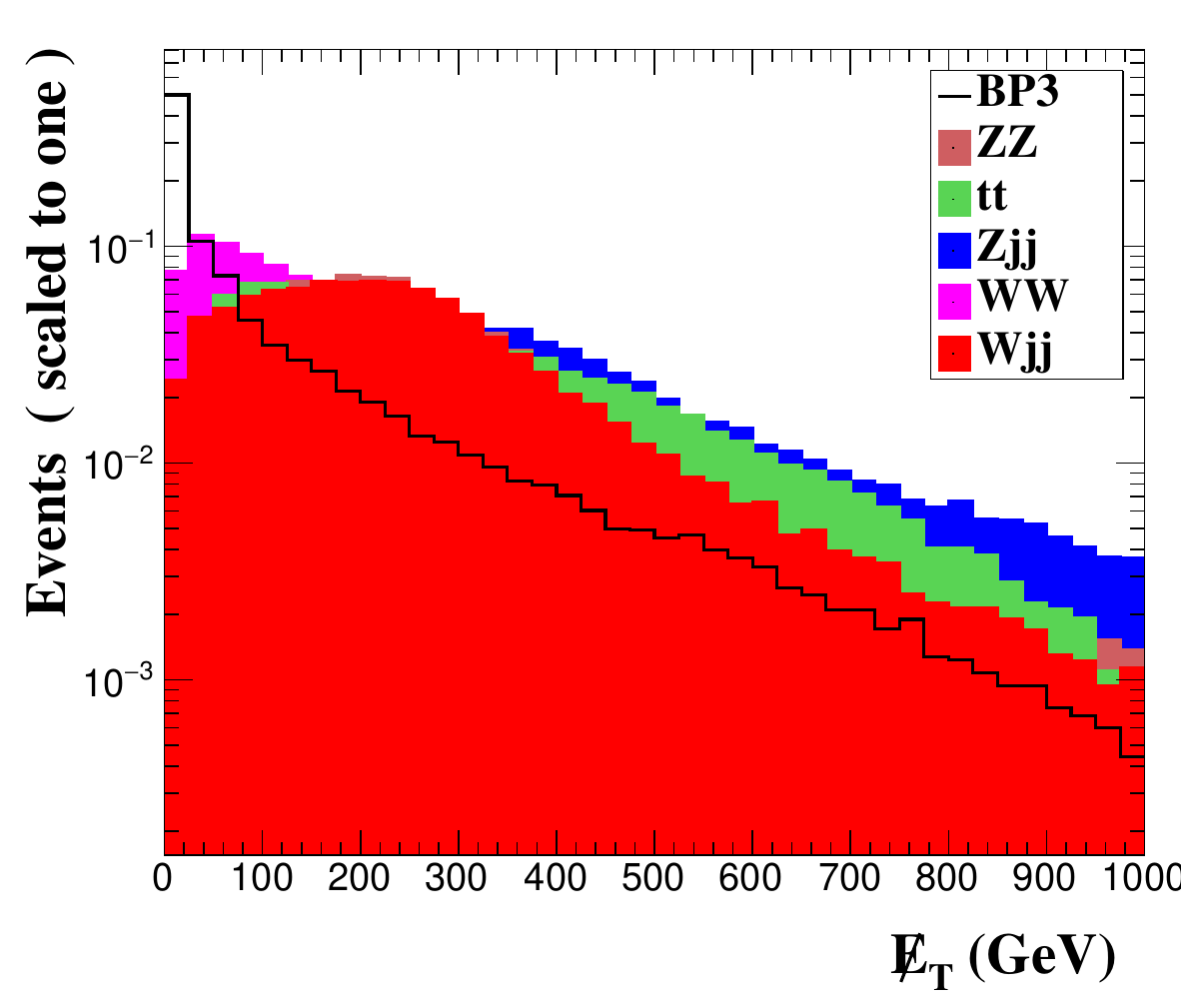}
\includegraphics[width=0.43\textwidth]{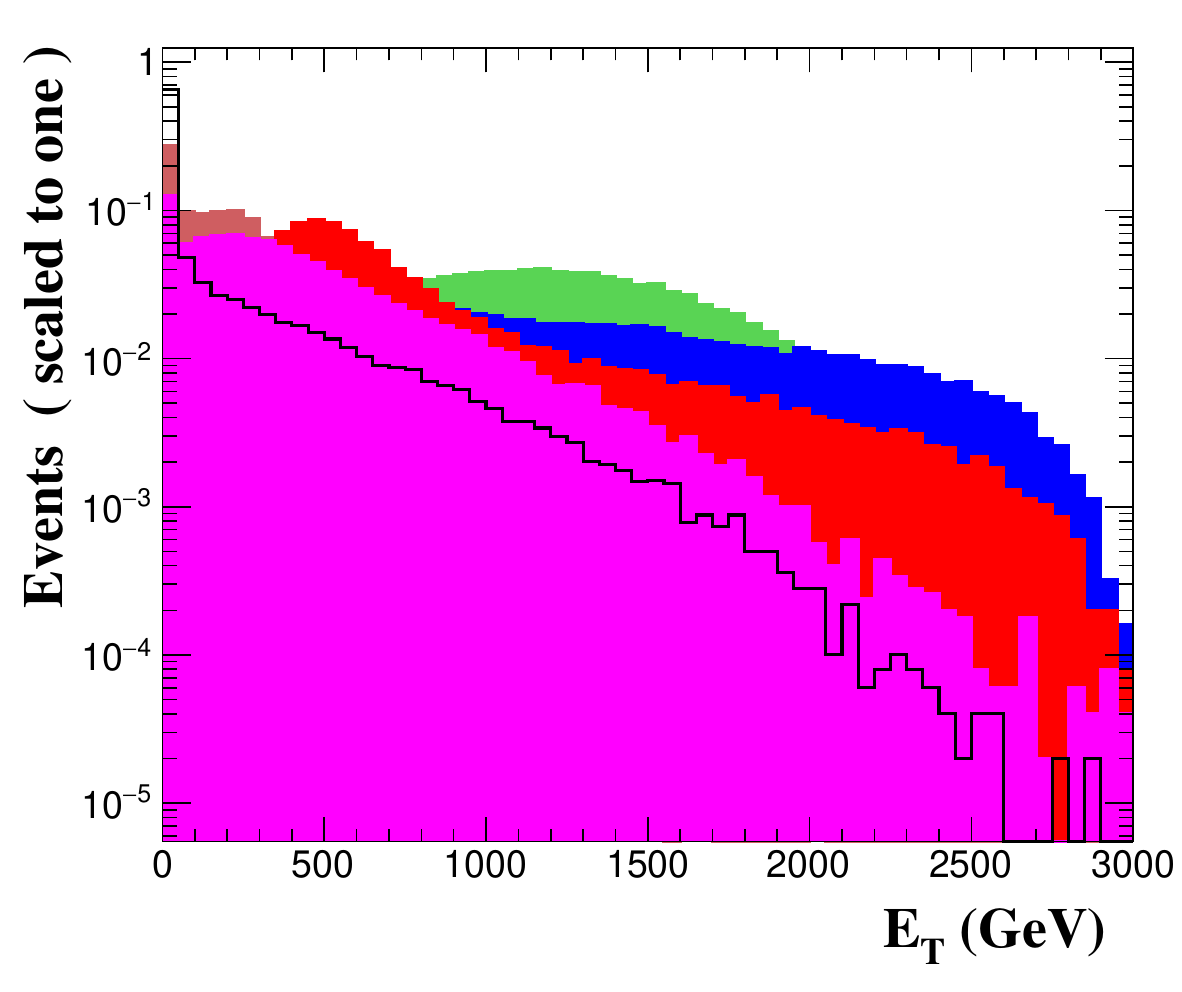}
\caption{Normalized kinematic distributions of the signal and backgrounds: the missing transverse energy $\slashed{E}_{T}$ (left panel) and the transverse energy $E_T$ (right panel) at $\sqrt{s}=$3 TeV muon collider.} \label{fig7}
\end{figure}
	
At parton level, the events of the signal and backgrounds are required to pass through the basic cuts as follows: 
\begin{equation} \nonumber
p_{T}^{j} > 20 \hspace*{4mm},\hspace*{4mm} \hspace*{4mm}  |\eta^{j}| < 2.5 ;.
\end{equation}
	
We show in Fig.\ref{fig7} the kinematic distributions of the signal and backgrounds for the missing momentum $\slashed{E}_{T}$ (upper-left panel) and  the transverse energy  $E_{T}$ (upper-right panel) at 3 TeV muon collider. To enhance the signal significance, we have implemented a cut-flow strategy guided by the kinematic distributions, outlined in Table \ref{cutft}. The initial step involves restricting the count of b-quarks, ensuring $N(b) \leq 1$. This constraint proves crucial in differentiating the signal from the background. With this cut, the $t\bar t$ background retains only 47\% of events, while minimally affecting the signal events.
As shown in the upper-left panel  $\slashed{E}_{T}$ plays a censorious role in separating the signal from background. So, we apply $\slashed{E}_{T} < 75$ GeV. Under this cut alone the surviving rate of the dominant backgrounds $Wjj$ is about $12\%$ and other backgrounds have already become very small, while the signal  survival rate is  about 67$\%$. We investigate one of the the most efficient cuts $E_{T} <$ 250 GeV, which keeps about 95$\%$ of the signal events and removes more than 92\% of $t\bar{t}$, $Zjj$, $WW$ and $Wjj$ background events. The selection cuts imposing are listed in Table~\ref{cuttts}.
	
	\begin{table}[h]
		\centering
		\setlength{\tabcolsep}{60pt}
		\renewcommand{\arraystretch}{1.2}
		\begin{adjustbox}{max width=\textwidth}
			\begin{tabular}{cc} 
				\hline  \hline
				Cuts & Definition \\  
				\hline \hline
				Trigger  & 	$N(b) < $  1 \\
				\hline
				Cut-1  &  $\slashed{E_{T}} <$ 75 GeV \\  
				\hline	
				Cut-2 & $E_{T} <$ 250 GeV  \\  		
				\hline \hline
			\end{tabular}
		\end{adjustbox}	
		\caption{A set of cuts used in the signal-background analysis
			for $\mu^+ \mu^- \to H^\pm W^\mp \to  \tau^+ \nu_{\tau} \tau^- \nu_{\tau}$
			at $\sqrt{s}=3$ TeV.}	\label{cuttts}
	\end{table}

	\begin{table}
		\centering
		\setlength{\tabcolsep}{5.pt}
		\renewcommand{\arraystretch}{1}
		\begin{tabular}{p{3cm}<{\centering}  p{1.2cm}<{\centering} p{1.4cm}<{\centering}p{1.4cm}<{\centering}  p{1.6cm}<{\centering} p{1.6cm}<{\centering} p{1.4cm}<{\centering} p{1.4cm}<{\centering} p{2cm}<{\centering} p{2cm}<{\centering}  p{2cm}<{\centering}p{0cm}<{\centering}}
			\hline\hline
			\multirow{2}{*}{Cuts$\ \ \ \ \ \ \ \ \ \ \ \ \ \ \ $}& \multicolumn{1}{c}{Signal }& \multicolumn{3}{c}{~~Backgrounds}&  \\ \cline{2-2}  \cline{4-8}
			&  $\text{BP3}$ &&$t\bar t $& $WW$  & $ZZ$ & $Wjj$&$Z/\gamma \ jj$\\
			\hline\hline
			Basic cut $\ \ \ \ \ \ \ \ $ &0.064 &&0.2 &1.54&0.014& 7.04& 0.74 \\
			Tagger $\ \ \ \ \ \ \ \ \ \ \ $& 0.064 &&0.09 &1.53 &0.014 &6.39 & 0.68\\
			Cut-1$\ \ \ \ \ \ \ \ \ \ \ \ \ \ $&0.043   && 0.01 &0.45 &0.0014&0.77 &0.073\\
			Cut-2$\ \ \ \ \ \ \ \ \ \ \ \ \ \ $&0.0416 && 0.0004&0.24 &0.0014 &0.068 &0.002\\
			Total efficiencies & $64\%$  && 0.2$\%$ & 15$\%$ & 10$\%$ & $9.6\%$ & $0.27\%$ \\
			\hline\hline
		\end{tabular}
		\caption{The cut flow of the cross sections (in fb) for the
			signal and SM backgrounds at $\sqrt{s}=3$ TeV muon collider
			with  our typical $\text{BP3}$\label{cut2}.}
	\end{table}

	\begin{figure}
		\centering
		\includegraphics[width=0.48\textwidth]{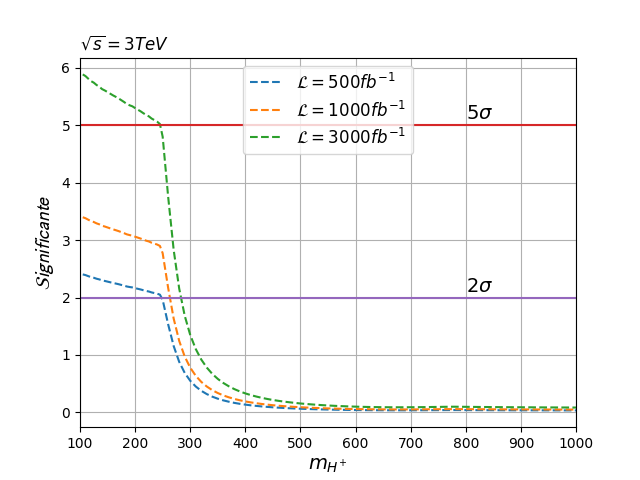}
		\caption{The significance obtained for the process $\mu^+ \mu^- \to H^\pm W^\mp \to \tau^+ \nu_{\tau} \tau^- \nu_{\tau}$ versus the charged-Higgs mass at 3 TeV muon collider with
			integrated luminosities of 500 fb$^{-1}$ , 1000 fb$^{-1}$ and 3000 fb$^{-1}$ with our benchmark point BP3.
			The 2$\sigma$ and 5$\sigma$ significance levels are also indicated.}
		\label{fig8}
	\end{figure}

	Fig.\ref{fig8} illustrates the 5$\sigma$ and 2$\sigma$ limit
	capability in the significance-$m_{H^+}$ plane at $\sqrt{s}=3$ TeV
	for various integrated luminosities. As shown, The 
	significance is presented as a function of $m_{H^+}$  for integrated luminosities of
	$\mathcal{L}=500 $fb$^{-1}$ , 1000 fb$^{-1}$ and 3000 fb$^{-1}$. We
	observe that a significance of 5$\sigma$  can be achieved when the integrated luminosity 3000fb$^{-1}$. The range of
	$m_{H^{+}}$ between  around 100 GeV and 250 GeV  is the most
	preferred region to discover the charged Higgs boson at
	$\sqrt{s}=3$ TeV muon collider. However, the 2$\sigma$ level achived within integrated luminosities 500fb$^{-1}$ and 1000 fb$^{-1}$. The integrated luminosities of
	$\mathcal{L}=500$ fb$^{-1}$, 1000 fb$^{-1}$ are significantly not very promissing. We find that under these integrated luminosities the significance becomes small, and the 5$\sigma$  level cannot be achieved under the available charged Higgs mass.
	In conclusion, it can be inferred that the values of integrated luminosities upper than 3000fb$^{-1}$ have a vital impact on achieving the 5$\sigma$ and
	2$\sigma$ significance. This also holds true for the quest to
	discover the charged Higgs boson $H^{\pm}$ through 
	the process $\mu^+ \mu^- \to H^\pm W^{\mp}$.

\section{Conclusions}
\label{sec:conclusion}
We have investigated the charged Higgs production at the upcoming muon
collider in the framework of 2HDM. We have studied
both $\mu^+ \mu^- \to H^+ H^-$ and $\mu^+ \mu^- \to W^\pm H^\mp$
and given the analytical amplitudes for
various Yukawa textures of the 2HDM.  The study was done taking into
account theoretical constraints as well as experimental ones such as
LHC Higgses searches as well as several B physics measurements.  We
demonstrated that one may obtain significant improvement for $\mu^+
\mu^-\to H^+ H^-$ compared to what we can obtain for $e^+ e^- \to H^+
H^-$, which is achieved by large $\tan\beta$ amplification
in the case of Type X. Such large $\tan\beta$ in 2HDM type X survive all 
kind of experimental or theoretical constraints.
The charged-Higgs boson can be probed through $\mu^+ \mu^-\to
H^+ H^-$ only for $m_{H^\pm}<\sqrt{s}/2$.  On the other side,
$\mu^+ \mu^- \to W^\pm H^\mp$ can be used to probe the charged Higgs
boson in the region $\sqrt{s}/2 \leq m_{H^\pm} \leq \sqrt{s}-m_W$, for which
the charged-Higgs pair production is not accessible. We have shown
that in this region one can still have a significant cross section for
$\mu^+ \mu^- \to W^\pm H^\mp$. \\
In the case of light charged Higgs boson mass, a significant enhancement in the production cross-section  Vector Boson Fusion (VBF) $e^+e^- \to \nu_e \bar{\nu}_e H^+ H^-$ may be obtained which makes this $2\to 4$ process compete with the other $2\to 2$.  
All the results presented for Type X are also valid for Type II
if we take into account constraints from B-physics that request that the charged Higgs
should be heavier that 680 GeV and well as $\tan\beta \leq 12$ as suggested by LHC Higgs data. However, once the charged Higgs boson is heavier than 680 GeV in Type II, the sensitivity is completely lost.
We have also performed a
signal-background analysis and obtained the discovery $5\sigma$ region
and the exclusion region $2\sigma$ at 3 TeV muon collider both for $\mu^+ \mu^-\to
H^+ H^-$,  $\mu^+ \mu^-\to
W^\pm H^\mp$ as well as Vector Boson Fusion (VBF).

\subsection*{Acknowledgments}
This work is supported by the Moroccan Ministry of Higher Education
and Scientific Research MESRSFC and CNRST: Projet PPR/2015/6. K.C. is
supported in part by the National Science and Technology Council of
Taiwan under the grant number MoST 110-2112-M-007-017-MY3.

\bibliographystyle{JHEP}
\bibliography{bibliography}
\end{document}